\documentclass[11pt,a4paper]{article}

\usepackage{amssymb}
\usepackage{amsmath}
\usepackage{bm}
\usepackage{graphicx}
\usepackage{siunitx}
\usepackage{booktabs}
\usepackage{multirow}
\usepackage{array}
\usepackage{xcolor}
\usepackage[margin=2.4cm]{geometry}
\usepackage{microtype}
\usepackage[numbers,sort&compress]{natbib}
\usepackage{authblk}
\usepackage{hyperref}
\usepackage[capitalise,nameinlink]{cleveref}
\usepackage{comment}
\usepackage{color}

\hypersetup{hidelinks}

\newcommand{\BTR}{BTR}
\newcommand{\RTB}{RTB}

\newcommand{\figbox}[1]{%
  \fbox{%
    \begin{minipage}[c][0.20\textheight][c]{0.92\linewidth}
    \centering
    \textbf{Figure placeholder}\\[2mm]
    #1
    \end{minipage}%
  }%
}

\begin{document}

\title{Booster-based beam recycling for swap-out injection at the High Energy Photon Source\thanks{Published in \textit{Accelerator: Science, Technology \& Applications} (2026), Article 100001. \url{https://doi.org/10.1016/j.accel.2026.100001}}}

\author[1,2]{Zhe Duan\thanks{Corresponding author: duanz@ihep.ac.cn}}

\author[1,2]{Jinhui Chen}

\author[1]{Yaoyao Du}

\author[1]{Yuanyuan Guo}

\author[1]{Jun He}

\author[1]{Xiyang Huang}

\author[1]{Daheng Ji}

\author[1,2]{Jingyi Li}

\author[1]{Fang Liu}

\author[1]{Peng Liu}

\author[1]{Zhi Liu}

\author[1]{Xiaohan Lu}

\author[1]{Yanhua Lu}

\author[1,2]{Cai Meng}

\author[1,2]{Yuemei Peng}

\author[1]{Saike Tian}

\author[1]{Guanwen Wang}

\author[1,2]{Jiuqing Wang}

\author[1,2]{Na Wang}

\author[1]{Yuanyuan Wei}

\author[1,2]{Gang Xu}

\author[1,2]{Haisheng Xu}

\author[1]{Yaliang Zhao}

\author[1]{Ying Zhao}

\author[1,2]{Yi Jiao\thanks{Corresponding author: jiaoyi@ihep.ac.cn}}

\author[1,2]{Weimin Pan\thanks{Corresponding author: panwm@ihep.ac.cn}}

\affil[1]{Institute of High Energy Physics, Chinese Academy of Sciences, Beijing 100049, China}
\affil[2]{University of Chinese Academy of Sciences, Beijing 100049, China}
\date{}

\maketitle

\begin{abstract}
Fourth-generation synchrotron light sources employ ultralow-emittance storage rings with stringent injection requirements.
On-axis swap-out injection alleviates the dependence on storage-ring dynamic aperture, but high-charge operation requires an efficient injector architecture capable of producing high-charge replacement bunches.
This paper presents the accelerator physics design and performance analysis of a booster-based beam-recycling swap-out injection scheme implemented at the High Energy Photon Source (HEPS).
In this approach, the full-energy booster serves as both an injector and a high-energy accumulator.
An extracted storage-ring bunch is returned to the booster, merged with a low-charge bunch previously injected from the linac and accelerated to full energy.
Following high-energy damping, the merged bunch is reinjected into the original storage-ring bucket.
The scheme avoids the need for a dedicated accumulator ring while enabling high-charge bunch replacement.
The recycling scheme was commissioned through staged machine studies.
Full recycling-chain simulations, commissioning studies, and measured performance analysis are presented.
The measured results characterize the recycling operation and quantify the transmission efficiency and performance limitations of the complete recycling loop.
These results demonstrate the feasibility of the booster-based beam-recycling architecture and establish its operational basis
for high-charge swap-out injection in future fourth-generation synchrotron light sources.

\end{abstract}

\medskip

\section{Introduction}
\label{sec:introduction}

Injection is a key design, technical and operational challenge for fourth-generation synchrotron light sources~\cite{chapman2023_fourthgen}. 
Their storage rings commonly employ multi-bend achromat~(MBA) lattices to reach ultralow natural emittance and high brightness~\cite{Eriksson2014,Tavares2018MAXIV,Liu2021Sirius,raimondi2023_ebs,Sajaev2025APSU,Aiba2026Crosstalk,Schroer2018PETRAIV}. 
The strong focusing and aggressive nonlinear optimization required by such lattices substantially reduce the available dynamic aperture, which can be only millimeter-scale at the injection point~\cite{Borland2014DLSR}. 
This makes conventional pulsed-orbit-bump off-axis accumulation increasingly difficult. 
Advanced injection schemes, such as pulsed-multipole injection~\cite{Harada2007PulsedQuadrupole,Takaki2010PSM,Leemann2012PulsedSextupole,Alexandre2021MIK,Ollier2023MIK}, 
longitudinal injection~\cite{Aiba2015LongitudinalInjection}, 
and anti-septum-based injection~\cite{Gough2017AntiSeptum}, can relax the dynamic-aperture constraint while retaining the capability of accumulating charge in the storage ring.

On-axis swap-out injection~\cite{Emery2003} provides a more radical solution by replacing stored bunches directly rather than accumulating injected charge in the storage ring.
In this scheme, a selected stored bunch or a stored bunch train is extracted from the storage ring and replaced by a corresponding on-axis injected bunch or bunch train.
Among injection schemes, it imposes the least demand on storage-ring dynamic aperture and therefore provides greater freedom for lattice optimization toward ultralow emittance and high brightness.
This approach introduces two principal challenges: realizing a fast kicker system with nanosecond-scale rise and fall times for selective bunch or bunch-train replacement, and producing high-charge replacement bunches with high transmission efficiency in the injector complex.
For high-charge swap-out injection, the latter requirement is often addressed by using a dedicated accumulator ring in the injector complex, as in the APS-U and ALS-U upgrade designs~\cite{fornek2019_apsu_fdr,harkay2019_par_highcharge,calvey:2025:napac,steier2018_alsu_concept,Ehrlichman2021ALSUKicker}.
Although effective, this route requires a separate accumulator ring and therefore increases the injector footprint, hardware scope, and commissioning complexity.

An alternative approach is to reuse the full-energy booster as a
high-energy accumulator that recaptures the bunch extracted
from the storage ring, merges it with a newly accelerated bunch,
and recycles the merged bunch to the same storage-ring RF
bucket after high-energy damping.
This booster-based beam-recycling architecture
~\cite{DuanIPAC2018} has been adopted for the High Energy
Photon Source (HEPS)~\cite{HEPSDesign,jiao2026_heps_commissioning} to enable high-charge
swap-out injection.
By recycling the extracted bunch rather than discarding it,
the amount of new charge that must be generated and accelerated
at low energy is substantially reduced, thereby relaxing the
demands on the injector front end while
avoiding the need for a dedicated full-energy accumulator ring,
together with its associated hardware, construction cost, and
operational complexity.

\begin{figure}[!htb]
\centering
\includegraphics[width=0.92\linewidth]{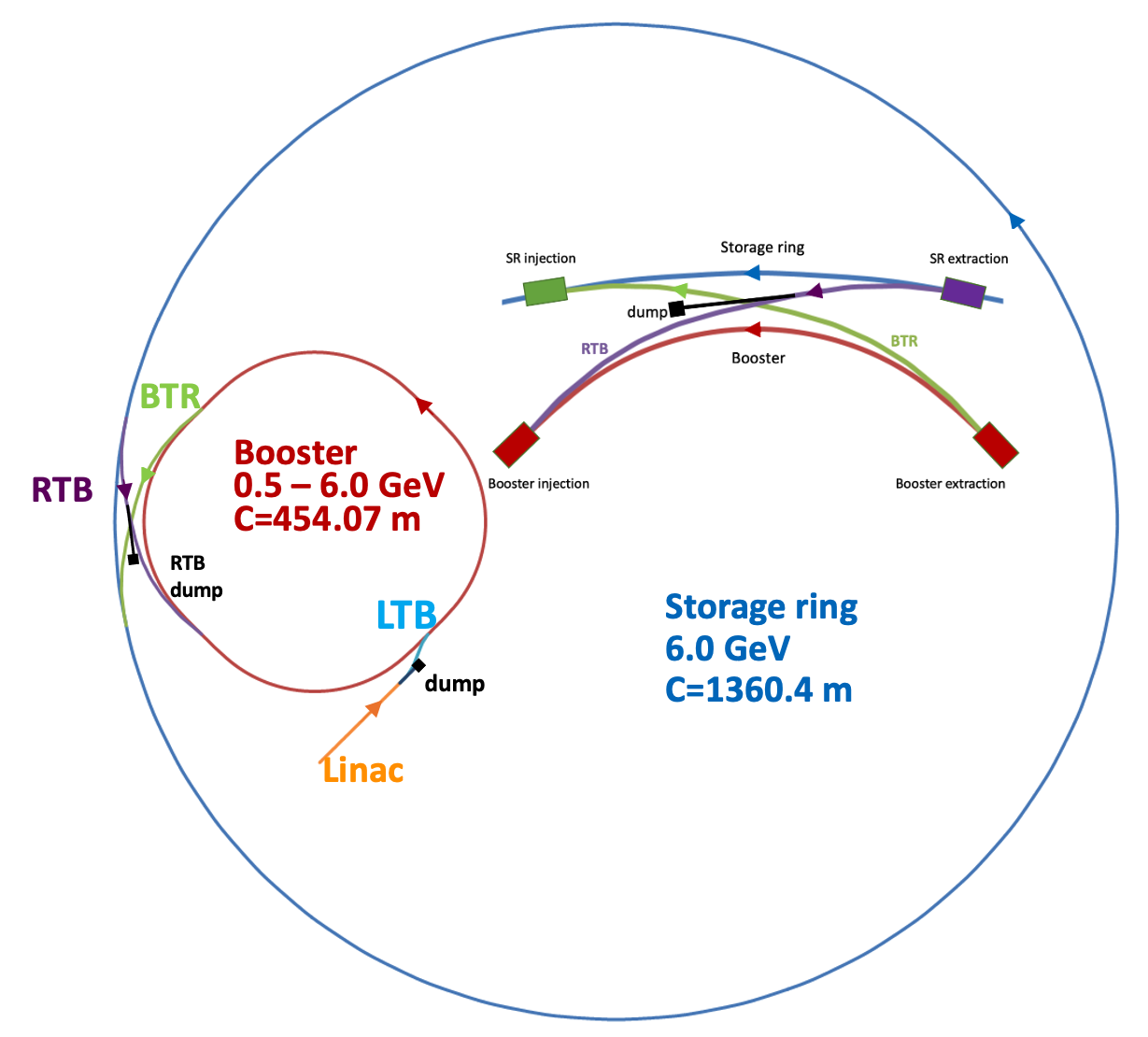}
\caption{Schematic layout of the HEPS injector complex and storage ring.
The enlarged inset shows the high-energy injection and extraction regions in the booster and storage ring.}
\label{fig:layout}
\end{figure}

HEPS is a \SI{6}{GeV} greenfield fourth-generation synchrotron light source based on a 48-cell 
seven-bend-achromat~(7BA) lattice with a design natural emittance of 
\SI{34.8}{pm$\cdot$rad}.
Its small dynamic aperture therefore requires on-axis swap-out injection as the baseline top-up method.
The HEPS injector complex consists of a \SI{500}{MeV} linac and a \SI{6}{GeV} full-energy booster.
As shown schematically in Fig.~\ref{fig:layout}, 
the full-energy booster together with the ring-to-booster~(RTB) and booster-to-ring~(BTR) transfer lines
provides the bidirectional high-energy transport path required for beam recycling.

Two representative operation modes were considered in the HEPS design stage~\cite{HEPSDesign}: 
a high-brightness mode and a high-charge timing mode. 
At \SI{200}{mA}, the high-brightness mode contains 680 bunches with a bunch charge of about \SI{1.33}{nC}, 
while the high-charge timing mode uses 63 bunches uniformly distributed around the storage ring, corresponding to a design bunch charge of about \SI{14.4}{nC}. 
The latter places particularly stringent demands on the injector system because of its large
single-bunch replacement-charge requirement.
The booster is designed to accelerate a \SI{5}{nC} bunch to \SI{6}{GeV}~\cite{peng2020_heps_booster}, 
enabling the required replacement charge to be achieved through repeated recycling cycles.
This greatly reduces the challenge of generating and accelerating high-charge bunches through the low-energy injector chain, where collective effects in the booster can become increasingly important at high bunch charge~\cite{foosang:2025:prst-ab}.

This paper presents the accelerator-physics basis, full recycling-chain simulation, staged commissioning, and performance decomposition of the booster-based beam-recycling scheme.
\Cref{sec:scheme} describes the overall beam-recycling sequence.
\Cref{sec:synchronization} presents the bucket-synchronization condition and timing implementation.
\Cref{sec:inj_ext_design} describes the storage-ring and booster injection/extraction region design and the associated aperture constraints.
\Cref{sec:simulation} describes the full recycling-chain simulation framework and the main results for both high-brightness and high-charge simulation cases.
\Cref{sec:commissioning} describes the staged commissioning process.
\Cref{sec:performance} presents the measured performance and the decomposition of the full-cycle efficiency.
\Cref{sec:conclusion} summarizes the main results and discusses the implications for future fourth-generation synchrotron light sources. %
\section{Overview of the beam-recycling swap-out injection scheme}
\label{sec:scheme}

\begin{figure}[!htb]
\centering
\includegraphics[width=0.92\linewidth]{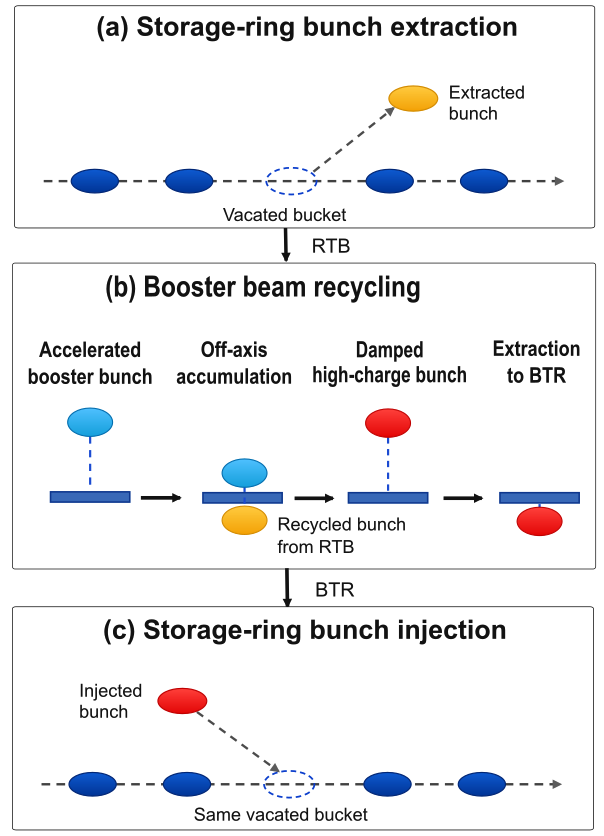}
\caption{Schematic of the beam-recycling swap-out injection sequence.
A selected storage-ring bunch is extracted from the storage ring, transported to the booster, merged with the low-charge bunch previously accelerated in the booster, damped at high energy, 
and returned to the same storage-ring bucket.}
\label{fig:sequence}
\end{figure}

The beam-recycling swap-out injection sequence is shown schematically in Fig.~\ref{fig:sequence}. 
A storage-ring bucket is first selected according to the filling pattern and bunch-charge target. 
A low-charge bunch from the linac is then injected into the booster and accelerated to \SI{6}{GeV}. 
The bunch originally stored in the selected bucket is extracted from the storage ring and transported through \RTB{} to the full-energy booster. 
It is injected into the selected booster bucket, where it is merged with the bunch already accelerated in the booster. 
After a high-energy damping interval, the merged bunch is extracted from the booster, transported through \BTR{}, and reinjected into the original storage-ring bucket.

The essential feature of this scheme is that the bunch extracted from the storage ring is reused rather than discarded.
The booster therefore serves a dual role: it accelerates the low-energy bunch injected from the linac to full energy
while simultaneously acting as a high-energy accumulator for
the bunch extracted from the storage ring.

The key parameters relevant to the beam-recycling scheme are summarized in \Cref{tab:parameters}. 
These parameters enter not only the same-bucket synchronization condition, but also the aperture analysis, beam transport, and simulation studies discussed in the following sections.
In addition, the storage ring employs a 166.6 MHz main RF system together
with 499.8 MHz harmonic cavities for bunch lengthening,
whereas the booster uses a 499.8 MHz RF system~\cite{zhang:2023:hepsrf}.
These RF configurations are relevant to the longitudinal capture and
phase-space evolution of the recycled bunch, as discussed in
\Cref{subsec:booster_longitudinal_capture}.

\begin{table}[!htb]
\centering
\caption{Key parameters relevant to the HEPS beam-recycling scheme.}
\label{tab:parameters}
\begin{tabular}{lll}
\toprule
Parameter & Symbol & Value \\
\midrule
Storage-ring energy & $E_{\rm SR}$ & \SI{6}{GeV} \\
Storage-ring circumference & $C_{\rm SR}$ & \SI{1360.39997}{m} \\
Storage-ring harmonic number & $h_{\rm SR}$ & 756 \\
Storage-ring natural emittance & $\epsilon_{\rm SR}$ & \SI{35}{pm$\cdot$rad} \\
Booster circumference & $C_{\rm B}$ & \SI{454.06648}{m} \\
Booster harmonic number & $h_{\rm B}$ & 757 \\
Booster natural emittance & $\epsilon_{\rm B}$ & \SI{16}{nm$\cdot$rad} \\
Storage-ring bucket spacing &  & $\simeq\SI{6}{ns}$ \\
Booster bucket spacing &  & $\simeq\SI{2}{ns}$ \\
Transfer-line length sum & $L_{\rm RTB}+L_{\rm BTR}$ & \SI{209.426}{m} \\
Coincidence-clock period & $T_{\rm cc}$ & \SI{3.435}{ms} \\
Coarse timing step &  & \SI{6}{ns} \\
Fine timing step &  & \SI{10}{ps} \\
\bottomrule
\end{tabular}
\end{table}

The recycling scheme is governed by four coupled requirements.
First, same-bucket synchronization requires the recycled bunch to return to the original storage-ring RF bucket.
Second, aperture and acceptance constraints determine the feasibility of extraction, transport, merging in the booster, and reinjection into the storage ring.
Third, high full-cycle transmission efficiency is required because repeated recycling cycles are needed in high-charge operation, and losses in either transfer direction 
reduce the net charge gain and operational margin. 
Fourth, operational reproducibility requires stable coordination among timing, kickers, diagnostics, and charge control.
These requirements motivate the synchronization, hardware, simulation, commissioning, and performance studies described in the following sections.
\section{Bucket synchronization and timing design}
\label{sec:synchronization}

\subsection{Same-bucket condition}
\label{subsec:same_bucket}

The synchronization design is implemented through the HEPS
timing system and injection-timing
framework~\cite{liu:2021:heps_gts}. The master oscillator (MO)
has a nominal frequency of \SI{166.6}{MHz}, which serves as the
reference for the storage-ring fundamental RF system, while its
third harmonic, nominally \SI{499.8}{MHz}, serves as the
reference for the booster RF system.

In the beam-recycling process, the bunch extracted from a
storage-ring bucket must be returned to the same bucket after
transport through the booster. This requirement can be expressed
as the path-length synchronization condition
\begin{equation}
M C_{\rm SR} - N C_{\rm B}
=
s_{\rm RE\rightarrow BI}
+s_{\rm BI\rightarrow BE}
+s_{\rm BE\rightarrow RI}
-s_{\rm RE\rightarrow RI},
\label{eq:synchronization}
\end{equation}
where $C_{\mathrm{SR}}$ and $C_{\mathrm{B}}$ are the
circumferences of the storage ring and booster, respectively.
The four $s$-terms denote the path lengths from storage-ring
extraction to booster injection, from booster injection to
booster extraction, from booster extraction to storage-ring
injection, and directly from storage-ring extraction to
storage-ring injection. The integers $M$ and $N$ are the numbers
of turns made by the empty storage-ring bucket and the recycled
bunch, respectively.

With the storage-ring and booster harmonic numbers
$h_{\mathrm{SR}}=756$ and $h_{\mathrm{B}}=757$, the synchronous
circumferences are related by
$C_{\mathrm{B}}/C_{\mathrm{SR}}=757/2268$.
Thus, for the nominal storage-ring circumference
$C_{\mathrm{SR}}=1360.39997~\mathrm{m}$,
the corresponding
synchronous booster circumference is
$C_{\mathrm{B}}=454.06648~\mathrm{m}$.
Because the two harmonic numbers are coprime,
Eq.~\eqref{eq:synchronization} admits discrete solutions,
$M = 502 + 757k$, $N = 1503 + 2268k$,
for which $M C_{\mathrm{SR}} - N C_{\mathrm{B}} \simeq 458.865~\mathrm{m} $.
This value defines the corresponding path-length constraint on
the right-hand side of Eq.~\eqref{eq:synchronization} and was
therefore incorporated into the design of the \RTB{} and \BTR{}
transfer lines. The design choice $k=4$, corresponding to
$M=3530$ and $N=10575$ and a booster dwell time of about
\SI{16}{ms}, is the first synchronization-compatible solution
that provides the required damping time discussed in
\Cref{subsec:booster_inj_ext_design}.

The coprime harmonic numbers also make arbitrary bucket pairing possible. The storage-ring and booster zero buckets realign every
\(T_{\rm cc}=h_{\rm B} C_{\rm SR}/c \simeq \SI{3.435}{ms}\),
corresponding to a coincidence-clock frequency of approximately \SI{291.11}{Hz}.
Because an additional delay equal to an integer number of $T_{\rm cc}$ preserves the same storage-ring–booster bucket relation, it can be used to vary the booster dwell time without changing the selected bucket pair.
A coarse delay table with a \SI{6}{ns} step selects the required storage-ring and booster bucket pair, while fine delay tuning with a \SI{10}{ps} step is used to optimize the kicker timing.

Machine imperfections introduce path-length errors and hence
arrival-time errors. When the MO frequency is adjusted to
compensate for a storage-ring circumference error
$\Delta C_{\mathrm{SR}}$, the synchronous circumferences of both
rings change coherently according to the RF-frequency relation
given above, resulting in a corresponding change in the
left-hand side of Eq.~\eqref{eq:synchronization}.
For $\Delta C_{\mathrm{SR}}=\pm2~\mathrm{cm}$, this contribution
is about $\pm0.7~\mathrm{cm}$.
 Assuming an additional $\pm2~\mathrm{cm}$ path-length variation from transfer-line and alignment errors, the total path-length variation is below about $\pm3~\mathrm{cm}$, corresponding to
$|\Delta t_{\mathrm{arrival}}| < 100~\mathrm{ps}$.
This scale is used as the practical arrival-time tolerance for recycling synchronization and is compared with the longitudinal capture margin in \Cref{subsec:booster_longitudinal_capture}.

\subsection{Timing sequences for the two transfer directions}
\label{subsec:timing_sequences}

The two high-energy transfer directions are implemented as separate event sequences within the same coincidence-clock framework. In the \RTB{} direction, the relevant triggers are the storage-ring pre-extraction kicker, the storage-ring extraction kickers, \RTB{} diagnostics, and the booster high-energy injection kickers. After the selected bunch is extracted from the storage ring, the booster injection kickers are triggered after the \RTB{} time of flight to recapture it in the selected booster bucket at \SI{6}{GeV}.

In the \BTR{} direction, the relevant triggers are the booster extraction bump magnets, the booster extraction kicker, \BTR{} diagnostics, and the storage-ring injection kickers. The four pulsed bump magnets are triggered about \SI{0.5}{ms} before extraction so that the local bump is fully established when the selected bunch reaches the extraction septum. The booster extraction kicker then sends the bunch into the \BTR{} line, and the storage-ring injection kickers are triggered after the \BTR{} time of flight.

In the injection-control application, the \RTB{} and \BTR{} event sequences can be selected and tuned separately. A complete recycling cycle is formed by selecting both sequences in the same booster cycle: after the booster reaches \SI{6}{GeV}, the \RTB{} sequence is executed first, and the \BTR{} sequence is executed later after a controlled delay. This delay determines the booster dwell time of the recycled bunch while preserving the bucket relation defined by the synchronization condition.

\section{Injection and extraction design}
\label{sec:inj_ext_design}

\subsection{Storage-ring injection and extraction regions}
\label{subsec:sr_inj_ext_design}

The storage-ring injection and extraction regions define the boundary conditions for the beam-recycling swap-out process. Since a single straight section of the HEPS storage ring is only about $6~\mathrm{m}$ long, it cannot accommodate both the injection and extraction hardware. 
The two regions are therefore placed in different straight
sections and adopt mirror-symmetric layouts,
with corresponding
stripline kicker modules and Lambertson septa adapted to the
opposite beam deflection directions. 
In the following, the injection region is described as the representative case; the extraction region follows the mirror-symmetric arrangement.

\begin{figure}[t]
\centering
\includegraphics[width=0.95\linewidth]{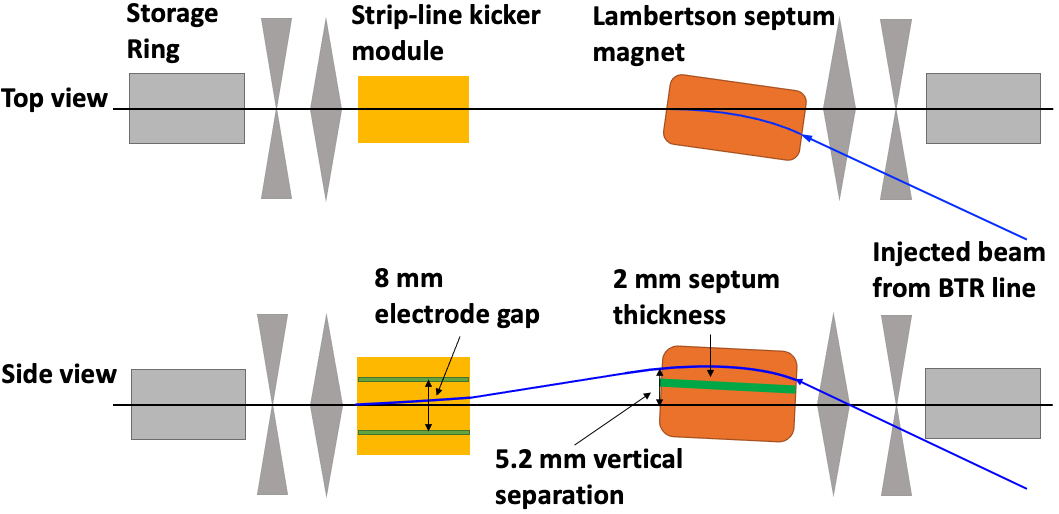}
\caption{
Schematic layout of the storage-ring injection region in top and side views.
The blue curve shows the injected-beam trajectory from the booster-to-ring transfer line, and the black line shows the stored-beam trajectory.
The Lambertson septum magnet merges the two trajectories, while the stripline kicker module provides the fast vertical kick required for on-axis swap-out injection.
The green strip denotes the septum blade.
The extraction region follows the mirror-symmetric layout.
}
\label{fig:sr_inj_ext_layout}
\end{figure}

In the injection region, as shown in Fig.~\ref{fig:sr_inj_ext_layout}, the Lambertson septum provides the main
separation between the stored-beam orbit and the injected-beam trajectory
from the BTR transfer line, while a compact stripline kicker module provides
the remaining fast vertical deflection. 
The geometry is determined by a compromise among septum thickness, stored-beam aperture, injected-beam clearance, kicker strength, pulse width, pulser voltage, and mechanical integration.
At the exit of the injection Lambertson, the vertical separation between the injected beam and the stored beam
$d_{\mathrm{sep}}$ is
$d_{\mathrm{sep}} = A_s + S + A_i = 5.2~\mathrm{mm}$, 
where $A_s=2.5~\mathrm{mm}$ is the stored-beam vertical clearance, $S=2.0~\mathrm{mm}$ is the effective septum thickness, and $A_i=0.7~\mathrm{mm}$ is the injected-beam clearance to the septum blade.
The Lambertson magnet provides a total bending angle of about $79.95~\mathrm{mrad}$ over an effective magnetic 
length of $1.6~\mathrm{m}$.
Because the Lambertson magnet is rolled with respect to the
storage-ring coordinate system, the deflection is primarily
horizontal with a small vertical component that contributes to
the required beam separation.

The stripline-kicker design is constrained by the bunch timing, requiring a sufficiently short kicker pulse for bunch-selective injection and extraction~\cite{Grishanov1997FastKicker}. In the high-brightness mode, the minimum bunch spacing is $6~\mathrm{ns}$; therefore, the effective kick-angle waveform must be confined to within about $12~\mathrm{ns}$ to avoid disturbing neighboring stored bunches. Since the high-voltage pulser already has a pulse base width of about $10~\mathrm{ns}$, the additional broadening from the stripline transit time limits the unit electrode length. For a stripline kicker of length $L_s$, the effective waveform is broadened by approximately $2L_s/c$, leading to the choice of $L_s=300~\mathrm{mm}$.

Shorter electrodes would further reduce the pulse width, but would require more stripline units and pulsers 
for the same kick strength. The adopted kicker module therefore contains five $300~\mathrm{mm}$ stripline 
kickers in a common vacuum chamber~\cite{wang2021_heps_5cell_kicker}, with an $8~\mathrm{mm}$ plate gap and a $6~\mathrm{mm}$ spacing 
between adjacent units. The required total vertical kick is $1.6108~\mathrm{mrad}$, corresponding to a high-voltage pulser specification within $\pm 16~\mathrm{kV}$ including correction factors, transmission losses, and operational margin. The kicker pulse is specified to be quasi-Gaussian, with a $3\%$–$3\%$ base width below $10~\mathrm{ns}$ and a residual amplitude below $3\%$ outside the main pulse.

The extraction region adopts the mirror-symmetric layout
described above. Besides the extraction kickers and Lambertson septum, a horizontal pre-extraction kicker is additionally included for machine protection. It is fired before the extraction kickers to dilute the selected stored bunch, reducing the maximum beam energy density on the extraction Lambertson septum in case of extraction-kicker failures while preserving normal extraction efficiency.

\subsection{Booster high-energy injection and extraction}
\label{subsec:booster_inj_ext_design}

\begin{figure}[p]
\centering
\includegraphics[width=0.95\linewidth]{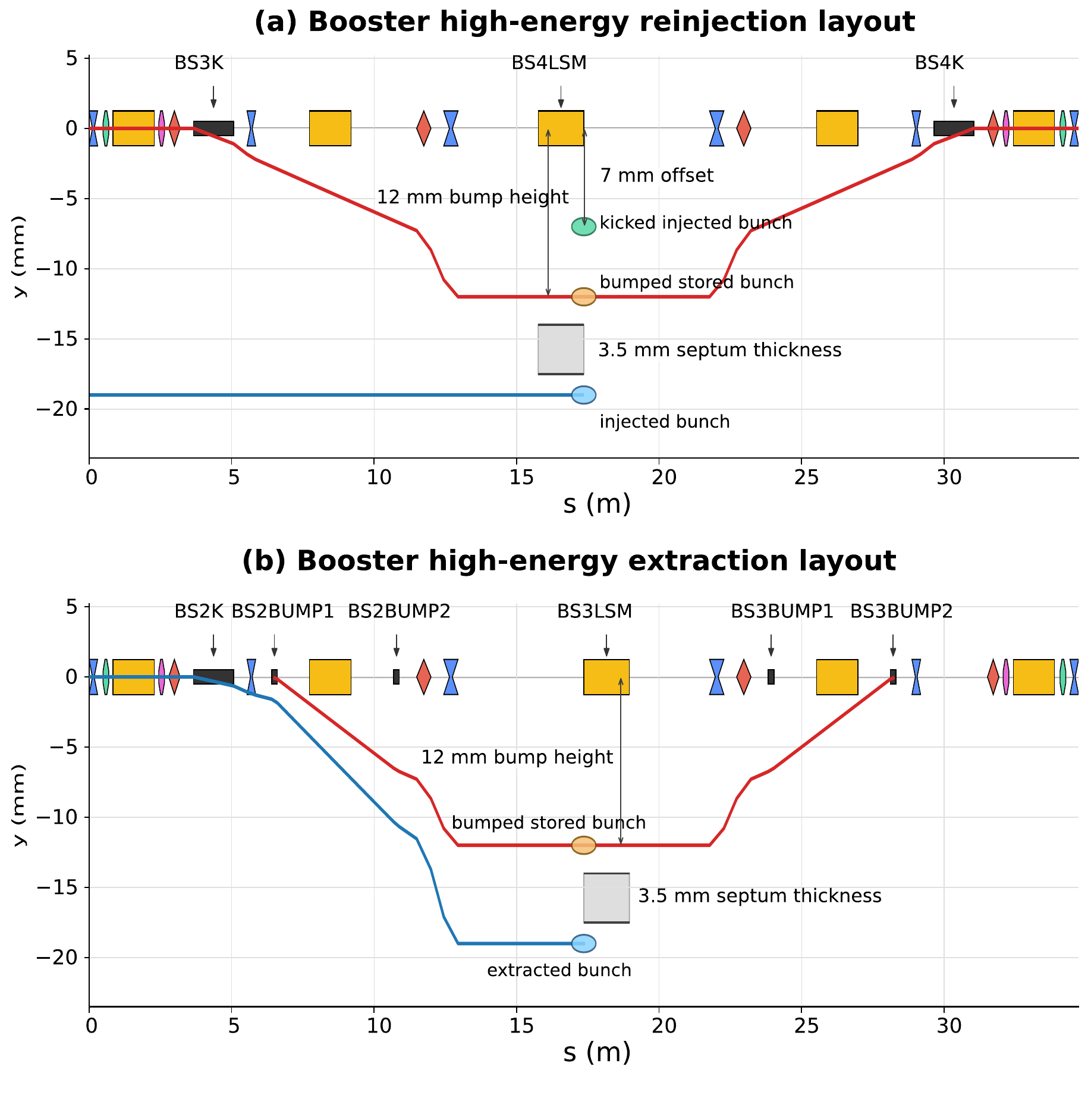}
\caption{Schematic layouts of the booster high-energy reinjection and extraction systems used in the beam-recycling process:
(a) high-energy reinjection from the storage ring to the booster and
(b) high-energy extraction from the booster to the storage ring.}
\label{fig:booster_inj_ext_layout}
\end{figure}

The booster high-energy injection and extraction layouts are shown schematically in Fig.~\ref{fig:booster_inj_ext_layout}.

Both booster high-energy injection and extraction use a Lambertson septum together with slotted-pipe kickers. For high-energy injection, two kickers generate a local vertical bump of about $12~\mathrm{mm}$ at the injection septum. The separation between the bumped circulating bunch and the incoming bunch is about $7~\mathrm{mm}$, consisting of a $2~\mathrm{mm}$ circulating-beam clearance, a $3.5~\mathrm{mm}$ septum thickness, and a $1.5~\mathrm{mm}$ incoming-beam clearance. The injection kickers act on the circulating bunch for one turn, with a pulse base width below \SI{1}{\micro\second}.

After high-energy injection, the recycled bunch has a residual vertical oscillation amplitude of about $7~\mathrm{mm}$. At \SI{6}{GeV}, the booster damping times are approximately $\tau_x=\tau_y=\SI{4.7}{ms}$ and $\tau_z=\SI{2.3}{ms}$. The bunch is therefore kept in the booster for at least about \SI{16}{ms}, corresponding to about 3.4 transverse damping times, so that the coherent oscillation and effective vertical emittance are sufficiently reduced
before extraction. In the design evaluation, the effective vertical emittance at storage-ring injection is reduced to within about \SI{4}{nm}, which is required to preserve the injected-beam clearance $A_i$ at the storage-ring Lambertson septum. The selected value $N=10575$ satisfies the same-bucket synchronization condition while providing this damping time.

For high-energy extraction, four pulsed bump magnets with a pulse width of about $1~\mathrm{ms}$ establish a local vertical bump before the extraction kicker is fired. 
The extraction kicker then deflects the selected bunch into the booster-to-ring transfer line. 
The extraction septum follows the same clearance concept as the injection septum, 
with a separation of approximately $7~\mathrm{mm}$ between the bumped circulating trajectory and the extracted-beam trajectory.

The booster high-energy extraction kicker has a required pulse width no longer than about $600~\mathrm{ns}$, 
so that the selected bunch can be extracted without significantly disturbing the other stored bunches.
 This pulse-width requirement limits the maximum number of bunches stored in the booster to five in the present recycling scheme.
 The commissioning operation reported here, however, used only one stored booster bunch per booster ramping cycle.

The pulsed-element requirements are determined by the need to
provide a sufficiently uniform and reproducible kick to
all particles in the injected or extracted bunch.
The effective flat-top width of each pulse must cover the bunch
duration together with the expected kicker timing jitter.
Within this time window, both the flat-top nonuniformity and the
shot-to-shot amplitude repeatability must remain small enough to
preserve the injection or extraction efficiency.
For instance, the storage-ring kickers require
sub-percent amplitude repeatability and
 an effective flat top of \SI{1}{ns} with an amplitude uniformity
 of less than $\pm$\SI{1.8}{\percent}.
These hardware-level requirements define the waveform quality
and stability targets for the storage-ring stripline kickers,
booster high-energy kickers, and booster bump magnets.
They also provide the basis for the pulsed-element error model
used in the full recycling-chain simulations in \Cref{sec:simulation}.

\section{Full recycling-chain transmission simulation}
\label{sec:simulation}

\subsection{Simulation setup}
\label{subsec:workflow}

The full recycling-chain transmission was evaluated with a chained multi-section tracking model implemented in elegant/Pelegant~\cite{Elegant,Pelegant}. 
The model follows one recycled bunch through storage-ring pre-extraction preparation, storage-ring extraction, \RTB{} transport, 
booster high-energy injection, recapture and extraction, \BTR{} transport, and final storage-ring reinjection. At each interface, the output particle distribution from the upstream section was used as the input distribution of the downstream section after applying the required coordinate transformations, static trajectory offsets, and arrival-time adjustments. This procedure preserves the beam halo and phase-space correlations generated upstream.

The initial bunch was generated according to the pre-extraction equilibrium distribution in the storage ring. In the transverse planes, the distribution was initialized using the corrected storage-ring lattice. In the longitudinal plane, the harmonic cavity was set to the ideal bunch-lengthening condition, and the phase-space distortion induced by longitudinal impedance was included. The selected bunch was then acted on by the horizontal pre-extraction kicker and tracked for 250 storage-ring turns before extraction. 

The simulations used the V3.1 storage-ring, booster, \RTB{}, and \BTR{} lattice models.
The storage-ring and booster seeds included static magnet field and alignment errors followed by closed-orbit, trajectory, tune, and optics corrections.
The \RTB{} and \BTR{} seeds were generated consistently with the corresponding storage-ring and booster seeds, including transfer-line magnet errors,
global trajectory correction, and local trajectory adjustment at the line ends. Residual static trajectory offsets at the transfer-line exits were included to
represent the finite precision of operational trajectory optimization.
 The simulations
assumed the designed vacuum-aperture geometry and therefore did not include unexpected
hardware obstructions that were identiﬁed after the reported beam measurements.

The pulsed elements were modeled with the \texttt{BUMPER} element in \texttt{elegant}. 
Measured waveforms were used for the storage-ring stripline kickers when available, 
while approximate half-sine waveforms were used for the booster high-energy kickers and 
bump magnets to represent the measured pulse shape and flat-top behavior. 
Shot-to-shot amplitude repeatability and timing jitter were then applied to these waveforms, 
as summarized in \Cref{tab:jitter_errors}. Amplitude errors are relative field errors for 
the \RTB{}/\BTR{} dipoles and quadrupoles, and relative voltage errors for the storage-ring and
booster kickers and bumpers. The listed values are treated as peak-to-peak engineering ranges sampled from uniform distributions, corresponding approximately to a conservative $4\sigma$ error envelope.

\begin{table}[!htb]
\centering
\caption{Main shot-to-shot jitter and repeatability errors used in the full recycling-chain simulations.
Amplitude repeatability errors are relative errors, and the listed values denote peak-to-peak engineering ranges sampled from uniform distributions.
}

\label{tab:jitter_errors}
\begin{tabular}{lll}
\hline
Error source & Error type & Error range \\
\hline
\RTB{}/\BTR{} dipoles
& Amplitude repeatability
& \(\pm 0.0002\) \\

\RTB{}/\BTR{} quads
& Amplitude repeatability
& \(\pm 0.00012\) \\

SR kickers
& Amplitude repeatability
& \(\pm 0.008\) \\

SR kickers
& Timing jitter
& \(\pm 200~\mathrm{ps}\) \\

Booster kickers
& Amplitude repeatability
& \(\pm 0.0048\) \\

Booster kickers
& Timing jitter
& \(\pm 5~\mathrm{ns}\) \\

Booster bumpers
& Amplitude repeatability
& \(\pm 0.0032\) \\

Booster bumpers
& Timing jitter
& \(\pm 10~\mathrm{ns}\) \\
\hline
\end{tabular}
\end{table}

\begin{figure}[p]
\centering
\includegraphics[width=\linewidth]{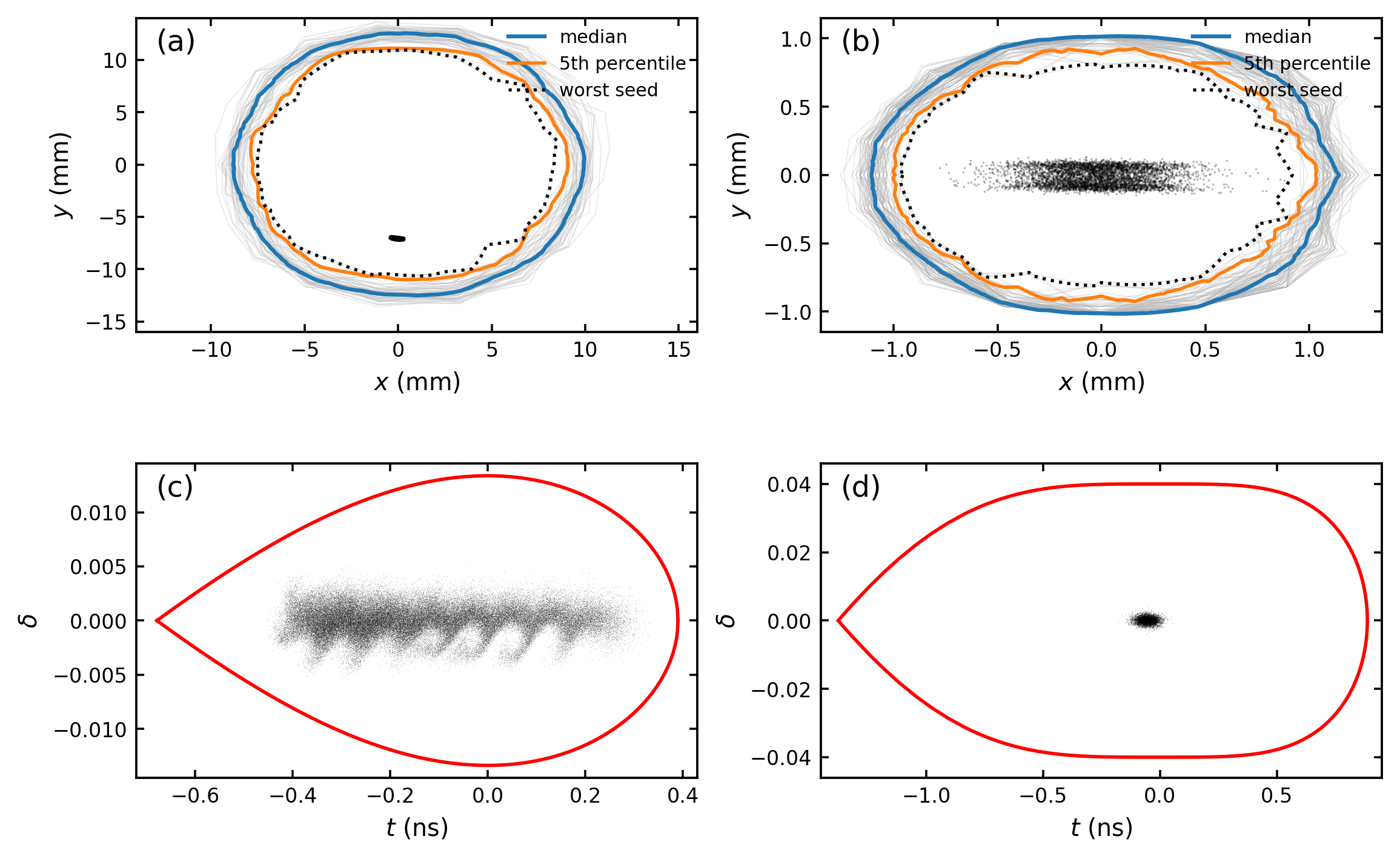}
\caption{
Comparison of tracked bunch distributions with the relevant transverse and longitudinal acceptances.
(a) Booster transverse dynamic acceptance at high-energy injection, with the injected beam placed at the nominal \SI{-7}{mm} vertical offset.
(b) Storage-ring transverse acceptance at reinjection, compared with a representative tracked bunch distribution from the high-charge full recycling-chain simulations after \BTR{} transport.
(c) Booster longitudinal RF acceptance at an RF voltage of \SI{8}{MV}, compared with the \SI{14.4}{nC} bunch distribution at booster high-energy injection.
(d) Storage-ring longitudinal acceptance, compared with the bunch distribution at storage-ring reinjection.
In (a) and (b), gray curves denote corrected imperfect-lattice dynamic-aperture boundaries, blue and orange denote the median and 5th-percentile boundaries, respectively, the black dotted curve denotes the most restrictive seed, and black points denote tracked particles.
}
\label{fig:acceptance_overview}
\end{figure}

For the high-brightness study, 50 corrected lattice seeds and 25 independent jitter seeds per lattice seed were used, giving 1250 complete full recycling-chain realizations. The extracted bunch was represented by 5000 macroparticles after the pre-extraction preparation. The booster tracking covered 10576 turns, corresponding to about \SI{16}{ms}, and the storage-ring reinjection was tracked for 2000 turns to evaluate transient capture loss. In this low-charge case, storage-ring impedance was included in the pre-extraction preparation, whereas impedance effects during booster tracking and final storage-ring injection were neglected to allow a broad statistical scan.

For the high-charge study, 20 corrected lattice seeds and 10 jitter seeds per lattice seed were used, giving 200 full recycling-chain realizations for the \SI{14.4}{nC} single-bunch study. 
Each bunch was represented by 100000 macroparticles. Vertical and longitudinal broadband impedance models were included in both the storage ring and the booster. 
These models account for the main single-bunch impedance-driven dynamics in this full recycling-chain transmission study, consistent with previous studies of collective effects and transient injection instabilities in ultralow-emittance storage rings~\cite{Sajaev2019Collective,Lindberg2019SwapOut}. A more complete treatment of high-charge transient collective effects, 
especially under large residual coherent oscillations during recapture and reinjection, is beyond the scope of the present full recycling-chain model.

The acceptance comparisons in Fig.~\ref{fig:acceptance_overview} were obtained from the same corrected lattices and tracked particle distributions used in the full recycling-chain simulations. 
In the transverse planes, dynamic-aperture boundaries from 50 corrected imperfect-lattice seeds were compared with tracked bunch distributions at the booster and storage-ring injection points. 
The booster-injection distribution includes the pre-extraction-kicker dilution and the subsequent \RTB{} transport, whereas the storage-ring injection distribution is a representative tracked case 
from the 200 high-charge full recycling-chain simulations after \BTR{} transport.
The transverse margin is large for booster high-energy injection, whereas storage-ring reinjection is more sensitive to the residual \BTR{} trajectory and injection-kicker errors.

\begin{table}[htbp]
\caption{
Statistical summary of the simulated full recycling-chain single-bunch transmission
efficiencies and stage-by-stage particle-loss distributions for the
high-brightness and high-charge simulation cases.
The particle-loss distributions are normalized to the total simulated
particle loss for each simulation case.
}
\label{tab:simulation_statistics}
\centering
\begin{tabular}{lcc}
\hline
 & High-brightness case & High-charge case (14.4 nC)\\
\hline

Number of realizations
&1250&200\\

Mean transmission efficiency
&99.99\%&98.97\%\\

RMS of transmission efficiency
&0.03\%&0.31\%\\

Minimum transmission efficiency
&99.68\%&98.12\%\\

\hline
\multicolumn{3}{c}{Stage-by-stage particle-loss distribution}\\
\hline

RTB transport
&0\%&0\%\\

Booster high-energy injection
&0\%&78.5\%\\

Booster extraction
&5.8\%&13.0\%\\

BTR transport
&73.2\%&1.3\%\\

Storage-ring injection
&21.0\%&7.2\%\\

\hline
\end{tabular}
\end{table}

For the longitudinal phase space, the \SI{14.4}{nC} bunch at booster high-energy injection was compared with the booster longitudinal acceptance, 
while the bunch at storage-ring reinjection was compared with the storage-ring longitudinal acceptance. 
The storage-ring harmonic cavity and longitudinal impedance significantly lengthen and distort the bunch before extraction, whereas the booster \SI{499.8}{MHz} RF system provides a relatively limited longitudinal acceptance. Booster high-energy longitudinal capture is therefore one of the tightest constraints in the high-charge case. By contrast, the reinjected bunch is shorter and well contained within the storage-ring longitudinal acceptance, so the storage-ring longitudinal bucket acceptance is not a dominant limitation.

Transmission was evaluated by counting surviving macroparticles at selected checkpoints, including the ends of the \RTB{} and \BTR{} lines, after booster capture and damping, and after storage-ring reinjection tracking. This simulation-based loss accounting is used to assess the design margin and identify dominant loss mechanisms.

\subsection{Full recycling-chain transmission results}
\label{subsec:end_to_end_results}

Table~\ref{tab:simulation_statistics} summarizes both the
overall transmission statistics and the stage-by-stage
particle-loss distributions for the simulated high-brightness and
high-charge cases.
The statistics are calculated over all lattice-error and jitter
realizations.
Because the overall particle loss is small, the stage-by-stage
loss fractions are reported to quantify how the residual losses
are distributed among the principal stages of the recycling
process. These fractions are obtained by accumulating the
particle losses at each stage over all realizations and
normalizing them to the total particle loss accumulated over the
corresponding simulation ensemble.

For the 1250 simulated high-brightness realizations, the mean overall transmission efficiency was
 99.99\%, and no realization exhibited a transmission efficiency below 99.6\%.
No particle loss was observed in the simulated
\RTB{} transport or at booster high-energy injection.
Only a small number of particles were lost in a small subset of
the simulated realizations during the booster extraction stage.
The remaining losses occurred predominantly in the downstream
\BTR{} transport and storage-ring injection stages.
The simulated \BTR{} losses were concentrated at the
storage-ring injection Lambertson septum, where the injected beam
passes a vertical one-sided physical aperture of about
\SI{0.7}{mm} relative to the injected-beam center.
These results demonstrate that the low-charge recycling process
has substantial operational margin under the modeled aperture,
correction, and jitter conditions.

The \SI{14.4}{nC} single-bunch study represents the most
stringent single-bunch transmission case.
No loss was observed during the \RTB{} transport.
The dominant loss mechanism occurred during booster
high-energy injection and the subsequent capture process,
where particles near the edge of the longitudinal phase-space
distribution exceeded the booster longitudinal acceptance,
as anticipated from Fig.~\ref{fig:acceptance_overview}(c).
Additional losses occurred during booster extraction in a subset of
lattice-error realizations, where particles trapped in
fifth-order resonance islands retained large vertical oscillation
amplitudes during damping and were scraped at the booster
extraction Lambertson as the pulsed extraction bump was ramped up.
Only a small fraction of the total loss occurred in the downstream
\BTR{} transport and storage-ring injection.
Consistent with the stage-by-stage loss statistics summarized in
Table~\ref{tab:simulation_statistics},
the modeled high-charge transmission is limited primarily by
booster longitudinal capture rather than by the downstream
transport.

The single-bunch transmission studies above do not include
the bunch merging process in the booster.
Charge accumulation was therefore investigated separately using
dedicated simulations for two representative high-charge accumulation scenarios.
At booster high-energy injection, the recycled storage-ring bunch
is merged with the circulating booster bunch, and the merged
distribution is subsequently tracked through the remainder of the cycle.
This treatment therefore includes the charge-dependent
impedance effects arising from the increased bunch charge after
the bunch merging and during the subsequent booster circulation.

For the representative top-up scenario,
the storage ring was assumed to operate with an average bunch-to-bunch
charge uniformity of about \SI{0.278}{\percent},
so that replenishing one bunch per top-up cycle requires an
average charge increase of approximately \SI{2.5}{nC}.
A representative depleted storage-ring bunch carrying
\SI{13.14}{nC}
was therefore merged with a
\SI{2.75}{nC}
circulating booster bunch.
To evaluate the charge gain under a representative operating condition rather than perform a statistical tolerance study,
the simulation was performed using one representative lattice-error
seed together with six independent jitter realizations.
After accounting for the simulated transmission losses throughout
the recycling cycle, the net charge gain was between
\SI{2.56}{nC} and \SI{2.66}{nC}.
This exceeds the required replenishment charge for maintaining
the nominal \SI{200}{mA} operation of the high-charge timing mode
under the assumed bunch-uniformity condition.
The result also indicates that bunch merging does not introduce
an additional dominant transmission limitation.

For filling from zero current, a representative charge-accumulation
sequence of
$0 \rightarrow 5 \rightarrow 10 \rightarrow 15~\mathrm{nC}$
was simulated for each storage-ring bucket, assuming that the
booster can accelerate a \SI{5}{nC} bunch to \SI{6}{GeV}.
Each accumulation step was evaluated using ten representative
lattice-error seeds with one independent jitter realization for
each seed.
The simulated charge after storage-ring capture reached
$9.996~\mathrm{nC}$
($9.992$--$9.999~\mathrm{nC}$)
for the
$5\rightarrow10~\mathrm{nC}$
step and
$14.984~\mathrm{nC}$
($14.972$--$14.991~\mathrm{nC}$)
for the
$10\rightarrow15~\mathrm{nC}$
step, where the values in parentheses denote the minimum and
maximum over the simulated realizations.
The corresponding mean overall transmission efficiencies 
were 99.964\% and 99.811\%, respectively,
consistent with the single-bunch transmission statistics in
Table~\ref{tab:simulation_statistics}.
These results indicate that, following the initial filling,
two additional recycling cycles per bucket are
sufficient to reach the nominal 14.4~nC bunch charge under
the simulated operating conditions, while explicitly accounting
for the bunch-merging process during booster high-energy
reinjection.

Overall, these accumulation simulations indicate that
the complete recycling process, including high-energy bunch
merging and repeated recycling cycles, can support both routine
top-up operation at the nominal 200~mA design point of the high-charge timing mode
and initial filling to the design 14.4~nC bunch charge under
the modeled operating conditions.

\subsection{Booster longitudinal capture during high-energy reinjection}
\label{subsec:booster_longitudinal_capture}

The booster longitudinal acceptance is one of the main constraints for high-charge recycling.
This limitation originates from the relatively large booster momentum
 compaction factor and energy loss per turn at \SI{6}{GeV},
 together with the choice of a \SI{499.8}{MHz} RF system.
 The implemented booster RF system uses six normal-conducting
 \SI{499.8}{MHz} cavities and provides a total voltage of
  about \SI{8}{MV} with the required operational margin.
  Further increasing the voltage would require additional cavities,
  which is constrained by the available length of the RF straight section and the overall booster layout. The present system therefore provides sufficient but limited longitudinal acceptance for high-energy capture.

This acceptance must contain the nominal \SI{14.4}{nC} design bunch used in the single-bunch transmission study.
 Before extraction, the storage-ring harmonic cavity lengthens the bunch, while longitudinal impedance further distorts its phase-space distribution. As a result, part of the longitudinal tail of the booster-injected bunch lies close to the booster RF acceptance boundary. 
Transient beam loading following the injection of a high-charge bunch can perturb the booster RF voltage and reduce the longitudinal capture margin, as previously studied for high-charge low-energy injection into the APS booster~\cite{Calvey2021APSBooster}.
The booster transient beam-loading effect during high-energy reinjection was modeled in elegant using the RFMODE element representing the \SI{499.8}{MHz} booster RF cavities. 
For the parameters studied, the additional efficiency reduction from this effect was below approximately \SI{1}{\percent}, and a suitable injected-bunch arrival-time window
 could recover a high capture efficiency.
 This indicates that booster transient beam loading contributes to the longitudinal capture margin, but does not dominate the modeled high-charge transmission loss under these conditions.

The injected-bunch arrival time is therefore an important matching parameter. Since the high-charge bunch occupies a large fraction of the booster RF bucket,
 a timing offset of order \SI{100}{ps} already produces a noticeable reduction in the capture margin, 
consistent with the path-length tolerance estimated from the synchronization condition in \Cref{subsec:same_bucket}. 
In operation, the arrival time can be tuned through the storage-ring RF phase.
The storage-ring harmonic-cavity settings provide an additional optimization knob: reducing the bunch-lengthening factor increases the booster capture margin, at the cost of storage-ring lifetime and IBS performance.
The present analysis therefore includes only the booster transient beam-loading effect relevant to high-energy recapture; storage-ring RF transients during the extraction--reinjection interval are left to a separate operational-impact study.

\subsection{Transmission constraints}
\label{subsec:transmission_bottlenecks}

The simulations indicate that the operational margin is mainly determined by three effects. The first is booster longitudinal capture at high-energy reinjection, including the finite booster RF acceptance, transient beam loading, and the injected-bunch arrival time.
The second is transverse halo generation and trapping in
fifth-order transverse resonance islands in a subset of booster
error seeds, primarily involving large vertical oscillation
amplitudes, which can lead to scraping at the booster extraction
Lambertson during the extraction-bump ramp.
The third is the tighter downstream acceptance of the \BTR{} and storage-ring injection sections, which makes the final capture sensitive to residual \BTR{} trajectory errors and storage-ring injection-kicker timing. Under the modeled assumptions, these effects do not prevent high-efficiency recycling, but they define the priorities for commissioning optimization.
The relative importance of these constraints depends on the
filling configuration, total beam current, and harmonic-cavity
condition, as discussed in \Cref{sec:performance}. %
\section{Staged commissioning}
\label{sec:commissioning}

\begin{figure}[htbp]
\centering
\includegraphics[width=\linewidth]{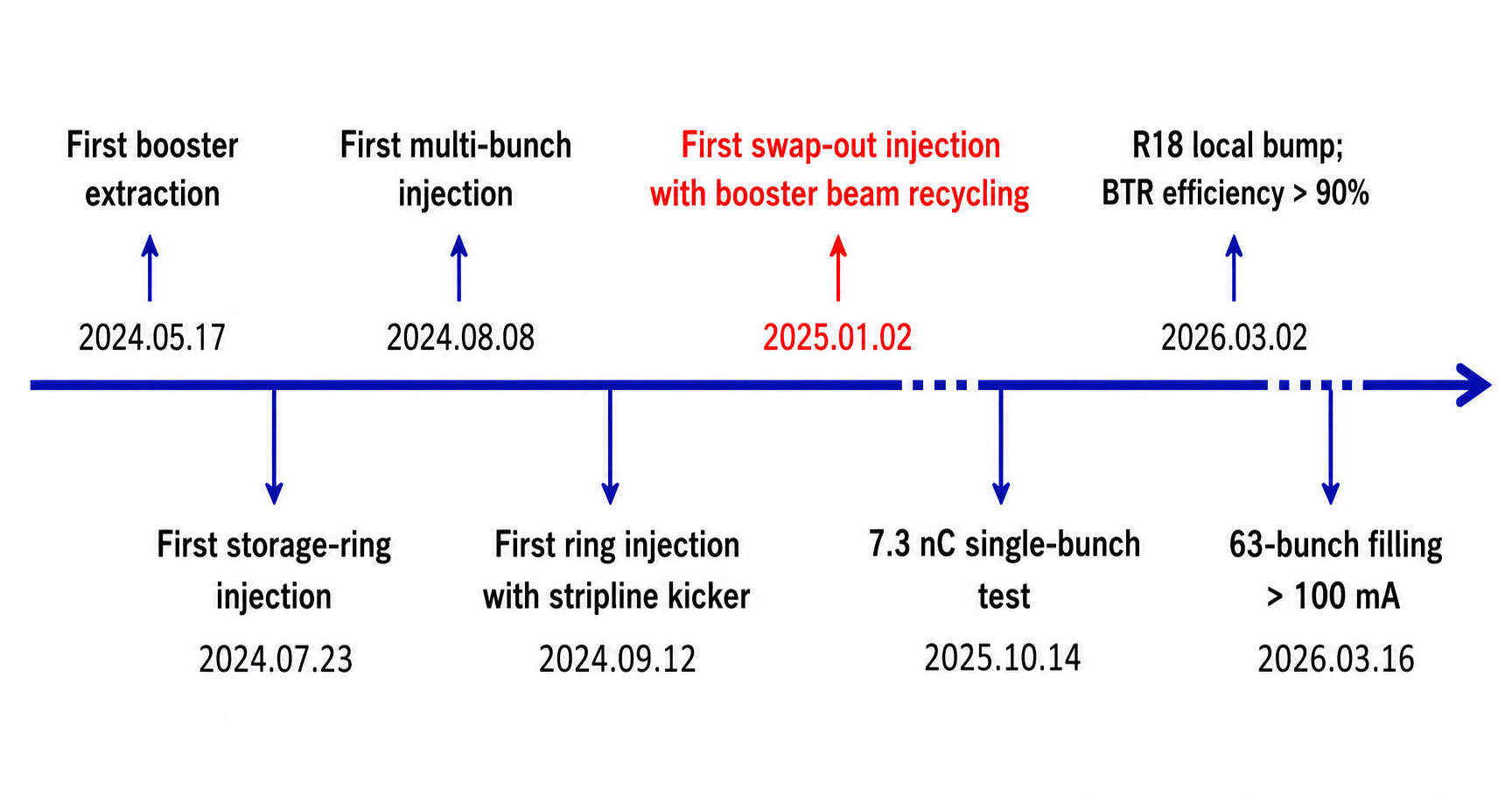}
\caption{Staged commissioning timeline for the high-energy injection/extraction and beam-recycling systems.}
\label{fig:commissioning_timeline}
\end{figure}

The booster-based recycling scheme was commissioned in stages to reduce the operational risk and to make each part of the full cycle independently verifiable.
The strategy was to first establish the booster-to-storage-ring direction,
then commission the storage-ring-to-booster direction,
and finally integrate the two event sequences in the same booster cycle according
to the same-bucket synchronization condition.
The main commissioning
milestones are summarized in Fig.~\ref{fig:commissioning_timeline},
including later high-charge optimization steps that mark the transition from proof-of-principle commissioning to routine operation.
Their quantitative performance is analyzed in \Cref{sec:performance}.

\subsection{Booster-to-ring transmission}
\label{subsec:btr_commissioning}

The first transfer direction to be established was the booster-to-ring path, including booster high-energy extraction, \BTR{} transport, and storage-ring injection. During this early stage, a temporary spare booster-type slotted-pipe kicker was used as the storage-ring injection kicker instead of the final stripline kicker. The temporary kicker provided a larger physical aperture, required only one pulser channel, and had already been tested in the booster high-energy extraction system, reducing the hardware and tuning risk for the first storage-ring injection tests. Its longer pulse width of about \SI{450}{ns} limited the available filling patterns, but this was acceptable for establishing stored beam and performing coarse orbit and optics corrections at a beam current of roughly \SI{10}{mA}.

Booster extraction required commissioning the local extraction bump, the extraction kicker, and the downstream trajectory in the \BTR{} line. A practical challenge was that the designed extraction bump amplitude exceeded the linear range of some booster beam position monitors~(BPMs). The bump response was therefore verified at reduced amplitudes within the linear BPM range and then scaled to the operational setting. Beam-based checks were also used to verify the polarity, relative response, and closure of the extraction bump magnets.

After booster high-energy extraction had been established, the \BTR{} trajectory was tuned toward the storage-ring injection point. The commissioning proceeded from transfer-line transmission to first-turn storage-ring capture, with beam-based trajectory correction performed using the \BTR{} BPMs and storage-ring first-turn diagnostics. The first storage-ring injection from the booster was achieved on 2024-07-23. Subsequent tuning established beam storage and then multi-bunch injection.

\subsection{Final stripline kicker validation and injection}
\label{subsec:stripline_commissioning}

While the temporary slotted-pipe kicker was used for early storage-ring injection, a final-type five-cell stripline kicker module was installed in the storage-ring extraction region for beam-based validation before the final injection stripline system was brought into operation. These tests covered possible beam-induced heating, the kick-strength response with high voltage applied to the stripline, and the timing-synchronization procedure for the multi-channel pulser system.

Each five-cell stripline kicker module is driven by ten high-voltage pulser channels. The relative timing of these channels must be synchronized to better than \SI{100}{ps} so that the individual kicks add coherently and provide the required total deflection angle. Insufficient channel synchronization would reduce the effective kick strength and introduce shot-to-shot kick jitter, directly degrading the extraction or injection efficiency.

Beam-induced signals on the stripline electrodes were used as an internal timing reference. The relative timing among the pulser channels was first adjusted using waveform measurements and then refined with beam response, so that the effective deflection of the selected bunch was maximized and made reproducible. After the temporary slotted-pipe kicker was replaced by the final injection stripline system, the same timing procedure was applied to storage-ring injection. Storage-ring injection using the final stripline kicker was achieved on 2024-09-12.

\subsection{Ring-to-booster transmission}

The opposite transfer direction required storage-ring extraction, \RTB{} transport, and high-energy reinjection into the booster.
After \SI{4}{ns} FID commercial pulsers became available,
they were installed to drive the storage-ring extraction kickers.
The storage-ring extraction timing was first established using beam-induced
signals and \RTB{} line diagnostics.
The extracted bunch was then transported through the \RTB{} line toward the booster high-energy injection point.

In the first high-energy reinjection tests, the extracted bunch arrived before the booster injection-kicker 
pulse became effective. 
After the event sequence was delayed by one booster turn,
scans of the delay and voltage of the downstream booster
reinjection kicker (BS4K), which provides the final kick for the
injected bunch while also acting on the circulating booster bunch,
established multi-turn survival,
and the booster DC current transformer~(DCCT) confirmed high-energy recapture of the 
extracted storage-ring bunch.
Based on the BS4K setting for injected-bunch capture,
the upstream booster reinjection kicker (BS3K), which acts
primarily on the circulating booster bunch, was then tuned
together with BS4K to form an approximately local orbit bump for the circulating \SI{6}{GeV} booster bunch. This minimized 
the disturbance to the circulating bunch while maintaining injection capture, and the booster DCCT 
confirmed high-energy accumulation of the recycled storage-ring bunch with the circulating booster bunch.

\subsection{Integration of the full recycling cycle}

After the two transfer directions had been commissioned separately,
they were combined into a single booster cycle.
The \BTR{} sequence defined the mapping from a selected booster bucket to a selected
storage-ring bucket, and the \RTB{} sequence was then adjusted to extract the same
storage-ring bucket and reinject it into the synchronized booster bucket.

In the integrated cycle, the booster is filled at low energy and ramped to \SI{6}{GeV}.
The selected storage-ring bunch is then extracted through the \RTB{} line
and recaptured in the booster, where it is accumulated and damped for a
controlled dwell time before being extracted through the \BTR{} line and
reinjected into the original storage-ring bucket.
During commissioning, the dwell time was adjusted by adding integer multiples of the coincidence-clock period, preserving the same-bucket condition while providing sufficient time for damping and charge diagnostics.

The first swap-out injection with booster beam recycling was demonstrated on 2025-01-02,
verifying that the two high-energy transfer directions could be coordinated as one closed same-bucket recycling cycle.

\subsection{Pre-extraction kicker and machine protection}

The integrated recycling sequence was later extended to include the pre-extraction kicker for machine protection,
and this configuration is used for routine operation.
Its timing was first adjusted using the beam-induced signal on the stripline electrodes, and the resulting excitation was then checked with turn-by-turn BPM data in the extraction region. With the optimized setting,
 the pre-extraction kicker was fired 250 turns before the main extraction sequence.
The observed residual horizontal oscillation was about \SI{0.3}{mm},
providing bunch dilution while remaining small compared with the downstream aperture
margin.
The subsequent extraction and \RTB{} reinjection sequence
was then shifted by 250 turns to include the pre-excitation in the full cycle.

\subsection{Beam-based optimization during commissioning}

Once the full recycling loop had been closed, the commissioning focus shifted from proof of sequence to transmission efficiency and charge scaling. The main transverse knobs were the transfer-line correctors, Lambertson settings, and kicker amplitudes and delays.
The practical objective was to minimize the maximum horizontal and vertical residual oscillations during the first tens to hundreds of turns after 
the two high-energy injection processes.

The longitudinal optimization mainly used two coupled tuning knobs: the booster extraction energy and the storage-ring RF phase. The booster extraction energy changes the relative energy mismatch in both transfer directions, and the storage-ring RF phase similarly changes the arrival-phase errors for both booster high-energy recapture and storage-ring reinjection. These knobs were therefore optimized to find an operational compromise between the \RTB{} and \BTR{} transmission efficiencies.

The available diagnostics were used to guide these beam-based optimizations.
 Turn-by-turn BPM data monitored residual oscillations after booster high-energy recapture 
 and storage-ring reinjection, while integrated current transformer~(ICT) and bunch charge monitor~(BCM) signals monitored relative charge transfer 
 during parameter scans. Beam loss monitor~(BLM) observations were used to identify beam-loss hot spots and guide 
 aperture-related tuning. Based on these beam-loss observations,
  a local orbit bump in the R18 arc region was introduced in March 2026 as an operational mitigation 
  to improve the storage-ring reinjection capture margin.
  The beam-loss observations that motivated this mitigation were 
  later found to be consistent with a localized aperture obstruction identiﬁed during the subsequent summer shutdown.
  The measured charge evolution and directional efficiencies obtained after this optimization are discussed in Section~\ref{sec:performance}.

\section{Measured performance and efficiency analysis}
\label{sec:performance}

The staged commissioning described in \Cref{sec:commissioning} established the 
feasibility of the full booster-based recycling cycle. 
This section summarizes the measured operational performance after the scheme had been integrated into 
routine operation. The emphasis is on charge evolution, directional transmission eﬃciencies, high-bunch-charge
ﬁlling capability, and the remaining optimization toward the nominal design point of the 
high-charge timing mode.

The measured efficiencies were obtained from charge diagnostics in the
single-booster-bunch commissioning configuration used for the present measurements.
For the storage ring, the total stored beam current was measured using the
storage-ring DCCT, while the BCM measured the relative bunch population of
each RF bucket. The BCM amplitudes were normalized to the corresponding
storage-ring DCCT current to obtain the absolute bunch charge. The storage-ring
BCM acquisition was synchronized with the RF system and updated at 10~Hz,
whereas the storage-ring DCCT provided the corresponding total-current
reference at 1~Hz. Since only a single bunch was stored in the booster during
the present measurements, the booster charge was determined directly from the
booster DCCT, which is updated at 10~Hz. The DCCT acquisitions were not directly synchronized to the
global timing system during these measurements. For the booster,
the energy ramp was used as a timing marker to identify the
DCCT readings before and after high-energy reinjection.

The performance of the charge diagnostics was independently characterized.
Both the storage-ring and booster DCCTs were calibrated against precision
current sources using linear calibration fits. The calibrated responses
exhibited residuals at the $10^{-5}$ and $10^{-4}$ levels relative to their
full-scale calibration ranges for the storage ring and booster, respectively.
The RMS resolution of the storage-ring DCCT was measured to be \SI{0.63}{\micro\ampere}
at 100~mA, while the booster DCCT had an RMS resolution of \SI{0.45}{\micro\ampere} at 5~mA.
The BCM
bunch-charge repeatability was better than 0.01~nC per bunch over 1000
consecutive measurements under a uniformly filled 252-bunch pattern at a total
stored current of 100~mA. 
The electronic pedestal and noise level of the BCM were
characterized using empty RF buckets, and bucket amplitudes
within the corresponding ADC noise range were set to zero
during digital processing.
Zero-current baselines were subtracted before DCCT charge analysis. The quantified diagnostic
uncertainties described above are much smaller than the measured charge
evolution and efficiency variations discussed below and therefore do not affect
the conclusions of the present analysis.

 These diagnostics provide the basis for a charge-based estimator of the directional transmission efficiency.
 For the \RTB{} direction, a charge-based estimator is
\begin{equation}
\eta_{\rm RTB}^{\rm meas}
=
\frac{Q_{\rm B}^{+}-Q_{\rm B}^{-}}{Q_{\rm SR}^{-}-Q_{\rm SR}^{+}},
\label{eq:eta_rtb_meas}
\end{equation}
where \(Q_{\rm SR}^{-}\) and \(Q_{\rm SR}^{+}\) are the selected storage-ring bucket charges before and after extraction, 
and \(Q_{\rm B}^{-}\) and \(Q_{\rm B}^{+}\) 
are the booster charges before and after high-energy reinjection, corrected for any pre-existing booster bunch charge. For the \BTR{} direction, a corresponding estimator is
\begin{equation}
\eta_{\rm BTR}^{\rm meas}
=
\frac{Q_{\rm SR,inj}^{+}-Q_{\rm SR,inj}^{-}}{Q_{\rm B,ext}^{-}-Q_{\rm B,ext}^{+}},
\label{eq:eta_btr_meas}
\end{equation}
where the numerator is the captured charge increase in the selected storage-ring bucket and the denominator is the charge extracted from the booster.
These estimators assume that the measured charge change during the analysis interval is dominated by the extraction and reinjection processes, while slow current decay and DCCT calibration uncertainties are negligible on this timescale.

The charge-based estimators were cross-checked with the turn-by-turn storage-ring BPM sum signals and supported by transfer-line ICT signals.
The BPM sum signals were used as an auxiliary indicator of storage-ring capture survival, while the ICTs provided relative charge-transfer information during trajectory and kicker scans. BLM signals were used qualitatively to identify loss locations, but were not used as absolute charge-loss monitors in the efficiency calculation.

\begin{figure}[htbp]
\centering
\includegraphics[width=\linewidth]{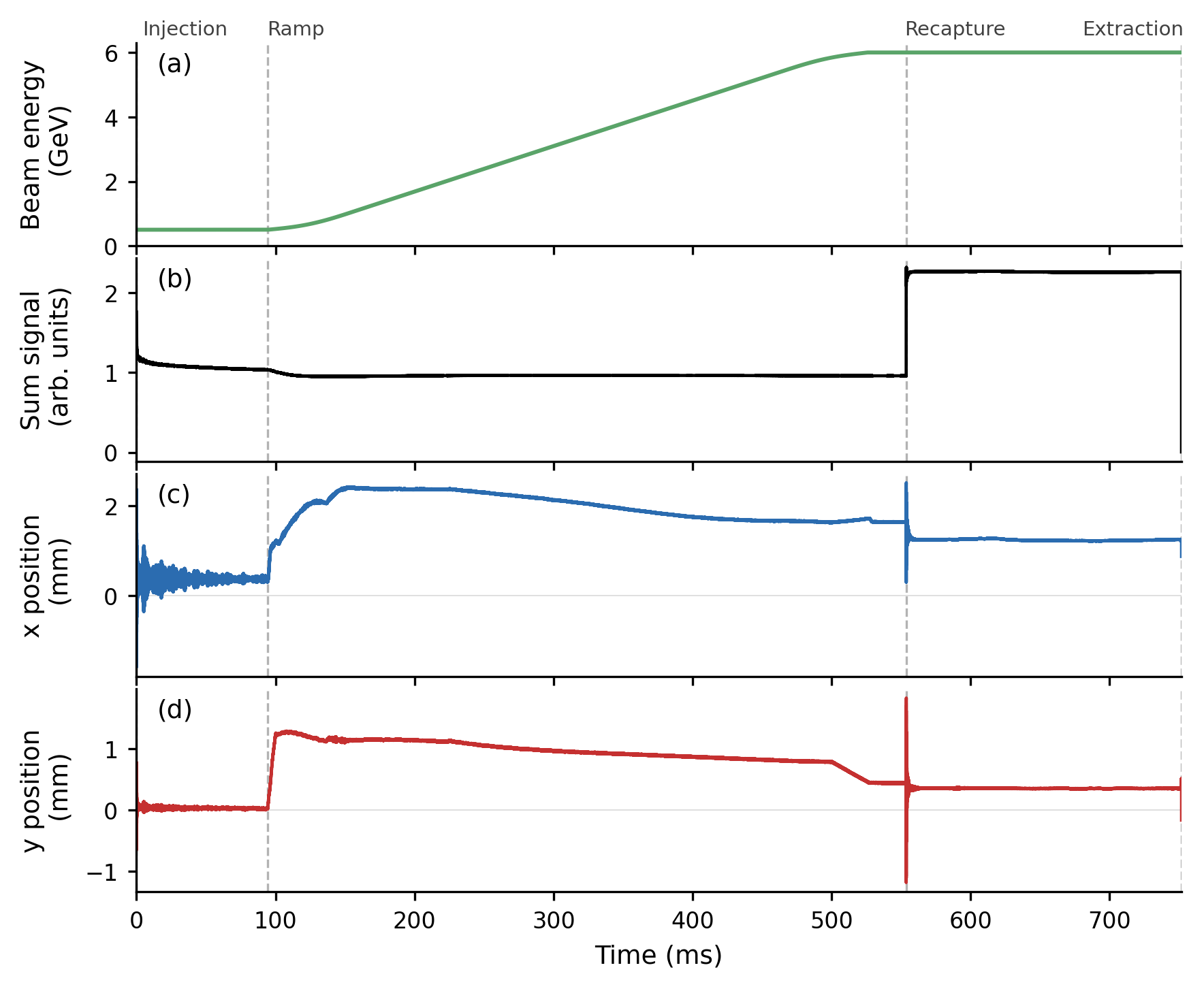}
\caption{Measured booster time trace during one recycling cycle.
The panels show the booster beam energy, bunch-sum signal, and horizontal and vertical turn-by-turn positions measured by a BPM.
The beam is injected at low energy, ramped to \SI{6}{GeV}, recaptured after high-energy reinjection from the storage ring, and finally extracted for storage-ring reinjection.}
\label{fig:single_cycle_booster_trace}
\end{figure}

A single-booster-bunch recycling cycle provides a compact time-domain validation of the sequence 
defined in \Cref{sec:synchronization}. Fig.~\ref{fig:single_cycle_booster_trace} 
illustrates the timing and beam-dynamics context of one recycling cycle using 
the booster time trace. The bunch is first injected into the booster at low 
energy and accelerated to \SI{6}{GeV} in about \SI{500}{ms}. The jump in the 
bunch-sum signal at high energy corresponds to recapture and accumulation of 
the extracted storage-ring bunch with the booster bunch. The horizontal and 
vertical turn-by-turn positions show the transient excitation after recapture. 
The trace also shows that the recycled 
bunch stayed in the booster for about \SI{220}{ms}, to facilitate the measurement of 
\(Q_{\rm B}^{-}\) and \(Q_{\rm B}^{+}\) with the \SI{10}{Hz} DCCT readout. 
This dwell time is much longer than the minimum value of about \SI{16}{ms} required for damping before storage-ring reinjection, and was used only for diagnostic convenience. 
It can be shortened when the booster DCCT update rate is increased to \SI{50}{Hz}.


 High-charge filling patterns require repeated recycling cycles
over many storage-ring buckets. The nominal high-charge timing
mode uses a uniformly distributed 63-bunch filling pattern, with
a design bunch charge of 14.4 nC at a total current of 200 mA.
In February 2026, dedicated operation with this 63-bunch filling
pattern was performed at 58--65 mA, corresponding to about
4.5 nC per bunch. 
After subsequent transmission-efficiency optimization, including
the use of a local orbit bump to improve the storage-ring
reinjection capture margin, the same 63-bunch filling pattern
reached \SI{101}{mA} on 2026-03-16.
This corresponds to an average bunch charge of about
\SI{7.3}{nC}.

The optimization procedure depended strongly on the filling
configuration and total beam current. For operation with the
63-bunch filling pattern, the higher total current required bunch
lengthening with the storage-ring harmonic cavity to mitigate
impedance-heating risks.
 This increased the longitudinal phase-space area of the bunch extracted for recycling and could reduce the booster high-energy capture margin. Optimizing the arrival time at booster reinjection then requires a coordinated phase adjustment of the \SI{166.6}{MHz} main RF and the \SI{499.8}{MHz} harmonic RF, so that the bunch-lengthening condition is maintained while the arrival phase with respect to the booster RF bucket is varied. This coupled optimization remains part of the ongoing high-charge commissioning.

By contrast, the 28-bunch high-charge test configuration had a lower total current, which allowed the harmonic cavity to be switched off and detuned. The extracted bunch was then lengthened mainly by impedance effects, 
and the booster longitudinal acceptance was sufficient for high-energy recapture in the present measurements. 
The 28-bunch data therefore provide a diagnostically cleaner measurement of the charge dependence of 
the recycling efficiency: this configuration partially decouples the transverse reinjection problem from 
the longitudinal matching problem. 
Operation at the nominal 200 mA design point of the high-charge
timing mode will still require both effects to be optimized
together under harmonic-cavity operation.

\begin{figure}[htbp]
\centering
\begin{minipage}{0.49\textwidth}
\centering
\includegraphics[width=\linewidth]{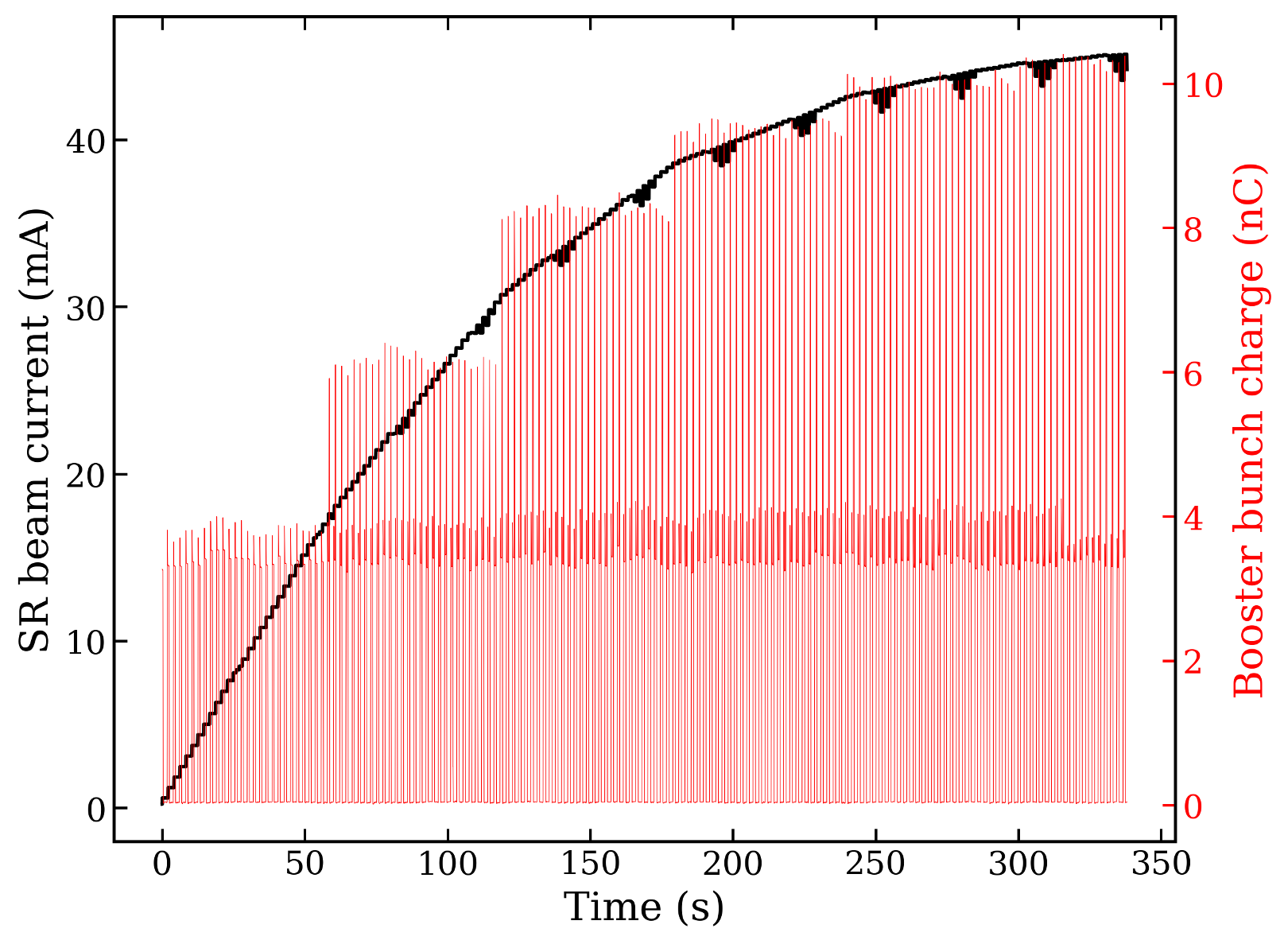}
\end{minipage}
\hfill
\begin{minipage}{0.49\textwidth}
\centering
\includegraphics[width=\linewidth]{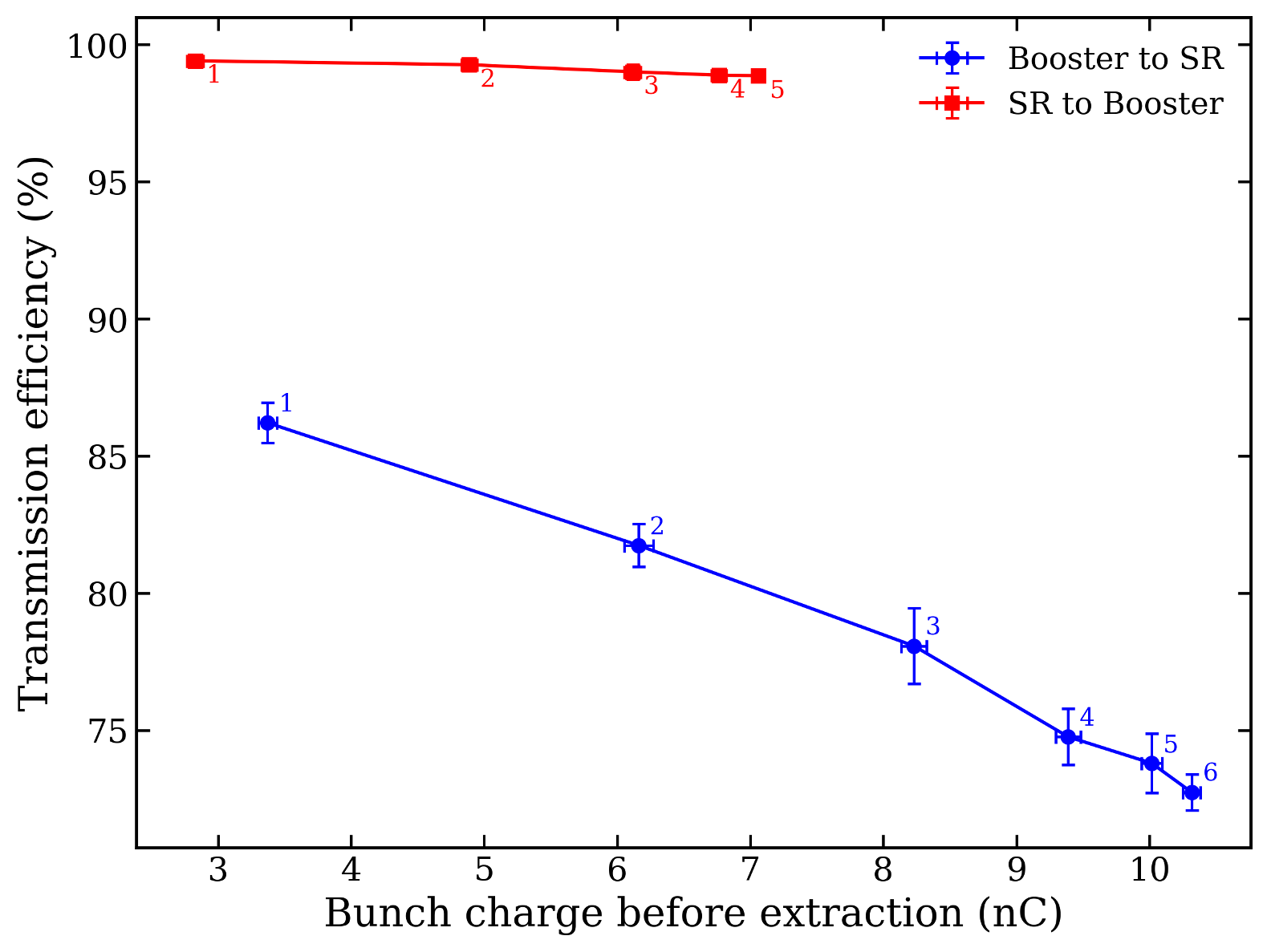}
\end{minipage}
\caption{Repeated filling and charge-dependent transmission in a 28-bunch high-charge filling test.
Left: storage-ring current accumulation and booster bunch charge during repeated recycling.
Right: measured directional transmission efficiencies as a function of the booster bunch charge before extraction.
The storage-ring-to-booster transmission remains close to \SI{99}{\percent}, whereas the booster-to-storage-ring efficiency decreases with increasing bunch charge, 
indicating that capture after storage-ring reinjection is the dominant measured limitation under the present operating conditions.}
\label{fig:repeated_filling_efficiency}
\end{figure}

To quantify this charge dependence under cleaner longitudinal conditions, a dedicated 28-bunch high-charge filling test was analyzed, as shown in Fig.~\ref{fig:repeated_filling_efficiency}. 
The 28 selected storage-ring buckets were filled sequentially according to bucket number. 
As a result, the booster bunch charge before extraction increased in a step-like manner, with each charge plateau corresponding approximately to one complete pass over the selected buckets. 
For each plateau, the \RTB{} and \BTR{} efficiencies were evaluated cycle by cycle using Eqs.~\eqref{eq:eta_rtb_meas} and \eqref{eq:eta_btr_meas}. 
The plateau-averaged efficiency is defined as the arithmetic mean of these cycle-by-cycle efficiencies over all recycling cycles belonging to the same charge plateau. 
The points in the right panel of Fig.~\ref{fig:repeated_filling_efficiency} show the plateau-averaged efficiencies, plotted against the mean booster bunch charge before extraction. 
The error bars represent the standard deviation of the corresponding cycle-by-cycle efficiencies within each plateau, but do not include possible common systematic uncertainties from the DCCT--BCM calibration.

As a reference for the booster-to-storage-ring direction, \BTR{} injection was also operated without booster high-energy reinjection, with the extracted storage-ring bunch dumped in the \RTB{} line. In this low-charge configuration, after optics correction and injection tuning, especially after applying the R18 local orbit bump, the routine \BTR{} injection efficiency could be stabilized above \SI{90}{\percent}. The lower \BTR{} efficiency of about \SIrange{85}{90}{\percent} at the first charge plateau in the 28-bunch recycling test should therefore not be interpreted as the optimized low-charge \BTR{} baseline, because the storage-ring injection trajectory in that test had already been tuned toward higher bunch charge.

An independent check was provided by the storage-ring BPM sum signals. For each injected bunch, the ratio of the BPM sum after 2000 turns to that on the first turn was used as an auxiliary capture-survival indicator. Its charge dependence was consistent with the \BTR{} efficiency derived from the DCCT/BCM estimator in Eq.~\eqref{eq:eta_btr_meas}, and no evidence was found for a separate charge-dependent loss in booster extraction or \BTR{} transport before first-turn capture.

The two transfer directions show clearly different behavior. 
The plateau-averaged storage-ring-to-booster efficiencies range from \SI{98.8}{\percent} to \SI{99.8}{\percent} 
over the measured charge range, indicating that storage-ring extraction, \RTB{} transport, 
and high-energy recapture in the booster have sufficient margin under these conditions. 
This high \RTB{} efficiency is consistent with the modeled \RTB{} transport and booster high-energy recapture process, 
and supports the basic soundness of the booster-based recycling concept. By contrast, 
the plateau-averaged \BTR{} efficiency decreases from \SI{86.2}{\percent} at the first charge 
plateau to \SI{72.7}{\percent} near \(Q_{\rm B,ext}^{-}\simeq\SI{10.3}{nC}\). Together with 
the low-charge reference operation and the BPM-sum cross-check described above, 
this indicates that the dominant measured limitation is the charge-dependent capture 
efficiency after reinjection into the storage ring, rather than booster extraction 
or transmission through the \BTR{} line itself.

This behavior is consistent with the different longitudinal conditions between the simulated design case and the present 28-bunch measurements. The simulations identify booster longitudinal capture as a tight constraint for the \SI{14.4}{nC} design bunch, especially when the storage-ring harmonic cavity lengthens the extracted bunch. In the 28-bunch high-charge test configuration,
the lower total current allowed the harmonic cavity to be switched off and detuned. The extracted bunch therefore had a smaller longitudinal phase-space area, and the booster capture margin was sufficient. 

The charge-dependent loss after \BTR{} transport is therefore not fully explained by the present full recycling-chain simulations, which predict high storage-ring capture efficiency for comparable bunch charge under the modeled aperture, trajectory, and jitter assumptions. 
This discrepancy suggests that the model captures the nominal lattice and jitter limitations, but not the full effective machine aperture and loss conditions during high-charge storage-ring reinjection. 
The improvement obtained with the R18 local orbit bump indicates that the effective storage-ring acceptance is influenced by localized loss conditions. 
Subsequent inspection of the vacuum chamber in the R18 local-bump region identiﬁed
a damaged ceramic cross support associated with a localized aperture obstruction near
the observed beam-loss hot spot. 
This finding provides a plausible explanation for the measured
charge-dependent loss: localized aperture restrictions may have
contributed to the observed efficiency degradation.
Their quantitative contribution remains to be verified after beam
operation resumes.

A significant charge-dependent transverse beam-size or mismatch increase during booster extraction, \BTR{} transport, or storage-ring reinjection is not expected from the present model, 
but separating any such contribution from aperture-related capture loss requires dedicated measurements. 
Existing bunch-resolved charge measurements, \BTR{} trajectory and kicker scans, and BPM-sum capture-survival checks provide the basis for the present efficiency analysis. 
Further optimization will focus on BLM-based loss localization, aperture analysis, charge-dependent effective-acceptance measurements, and injected-beam size and mismatch characterization. %

\section{Conclusion and outlook}
\label{sec:conclusion}

We have presented the design, simulation, staged commissioning, and measured performance of the booster-based beam-recycling swap-out injection scheme at HEPS. 
By exploiting the existing full-energy booster for high-energy recycling and damping, the scheme avoids the need for a dedicated accumulator ring while enabling same-bucket replacement of high-charge storage-ring bunches.

Full recycling-chain simulations indicate high transmission efficiency in both high-brightness and high-charge cases under realistic error and jitter assumptions. They also identify the main operational margins of the scheme: longitudinal capture at booster high-energy reinjection, halo generation and resonance-island trapping in some booster error seeds, and the downstream acceptance of the \BTR{} and storage-ring injection sections. These simulations provided the basis for the staged commissioning strategy and for interpreting the measured efficiency limitations.

Staged commissioning verified the two high-energy transfer directions separately before closing the full 
recycling loop. The \BTR{} process established booster-to-storage-ring transfer and storage-ring injection, 
while the \RTB{} process verified storage-ring extraction, ring-to-booster transport, and booster high-energy 
recapture. 
The two event sequences were then integrated within the same
booster cycle according to the same-bucket synchronization
condition.
This integration led to the first swap-out injection with booster
beam recycling.
Repeated filling experiments further demonstrated reproducible high-bunch-charge operation, 
including a uniformly distributed 63-bunch filling pattern above \SI{100}{mA},
representing an intermediate
step toward the nominal design point of the high-charge timing mode.

A central result of the measured performance is the high efficiency of the \RTB{} direction. 
This shows that the booster has sufficient high-energy acceptance to recapture the extracted 
storage-ring bunch and merge it with newly accelerated charge, which is a key validation of the 
booster-based recycling concept. The observed charge-dependent efficiency reduction appears instead 
in the booster-to-storage-ring direction. 
The available diagnostics indicate that, under the present
operating conditions, the dominant measured loss occurs after
BTR transport, during capture after reinjection into the storage ring.
No comparable dominant loss is observed during booster extraction
or transmission through the BTR line itself.
A localized aperture restriction in the R18 region, identiﬁed after the reported measure-
ments, may have contributed to the observed BTR eﬃciency degradation, although its
quantitative contribution has not yet been veriﬁed.

These results establish the feasibility of booster-based beam recycling as a practical route toward 
high-charge swap-out injection in future fourth-generation synchrotron light sources.
By exploiting the existing full-energy booster for high-energy recycling, this approach provides an alternative injector architecture to dedicated accumulator-ring solutions.
More broadly, the same-bucket booster-recycling principle may also be relevant to future circular colliders. For example, CEPC adopts swap-out injection in the Higgs-energy regime and considers a booster-recycling-based scheme for high-charge replacement bunches~\cite{CEPCAcceleratorTDR}.
The detailed collider-specific constraints are beyond the scope of this paper.

The remaining optimization of high-charge storage-ring capture after reinjection will determine the 
achievable bunch charge and operational margin of the HEPS high-charge timing mode relative to its design target.
Storage-ring RF transients during the extraction--reinjection interval are related to stored-beam perturbations and user-experiment stability and will be addressed in a dedicated study.
\section*{Declaration of generative AI and AI-assisted technologies in the manuscript preparation process}

During the preparation of this work, the authors used ChatGPT and DeepSeek for language editing, assistance with reference and LaTeX formatting, and assistance with figure formatting and visualization
of simulation and experimental data.
All simulation and experimental data used in the figures were generated by the authors. The authors reviewed and edited the output as needed and take full responsibility for the content of the published article.

\section*{Acknowledgments}

The authors thank the HEPS accelerator, injection/extraction, timing, RF, controls, diagnostics, magnet, power-supply, vacuum, and commissioning teams for their contributions to the design, commissioning, and operation of the beam-recycling swap-out injection system.
This work was supported by High Energy Photon Source (HEPS), a major national science and technology infrastructure, the National Natural Science Foundation of China (NSFC, 12275284, 12375149), the National Key R\&D Program of China (Grant No. 2022YFA1603401), and the Youth Innovation Promotion Association CAS (No. 2021012).

\section*{Data availability}

Data will be made available on request.

\bibliographystyle{unsrtnat}
\bibliography{references}

@inproceedings{Emery2003,
  author    = {L. Emery and M. Borland},
  title     = {Possible Long-Term Improvements to the {Advanced Photon Source}},
  booktitle = {Proceedings of {PAC}'03},
  address   = {Portland, OR, USA},
  year      = {2003},
  pages     = {256--258},
  paper     = {TOPA014},
  url       = {https://jacow.org/p03/papers/TOPA014.pdf}
}

@article{Eriksson2014,
  author = {Eriksson, Mikael and van der Veen, J. Friso and Quitmann, Christoph},
  title = {Diffraction-limited storage rings -- a window to the science of tomorrow},
  journal = {Journal of Synchrotron Radiation},
  volume = {21},
  pages = {837--842},
  year = {2014},
  doi = {10.1107/S1600577514019286}
}

@article{HEPSDesign,
  author = {Jiao, Yi and Xu, Gang and Cui, Xiao-Hao and Duan, Zhe and Guo, Yuanyuan and He, Ping and Ji, Daheng and Li, Jingyi and Li, Xiaoyu and Meng, Cai and Peng, Yuemei and Tian, Saike and Wang, Jiuqing and Wang, Na and Wei, Yuanyuan and Xu, Haisheng and Yan, Fang and Yu, Chenghui and Zhao, Yaliang and Qin, Qing},
  title = {The {HEPS} project},
  journal = {Journal of Synchrotron Radiation},
  volume = {25},
  pages = {1611--1618},
  year = {2018},
  doi = {10.1107/S1600577518012110}
}

@inproceedings{DuanIPAC2018,
  author = {Duan, Zhe and Chen, J. and Guo, Y. Y. and Jiao, Y. and Li, J. L. and Peng, Y. M. and Wang, J. Q. and Wang, N. and Xu, G. and Xu, H. S.},
  title = {The Swap-Out Injection Scheme for the {High Energy Photon Source}},
  booktitle = {Proceedings of IPAC'18},
  venue = {Vancouver, Canada},
  publisher = {JACoW Publishing, Geneva, Switzerland},
  pages = {4178--4181},
  year = {2018},
  doi = {10.18429/JACoW-IPAC2018-THPMF052},
}

@techreport{Elegant,
  author = {Borland, M.},
  title = {{ELEGANT}: A Flexible {SDDS}-Compliant Code for Accelerator Simulation},
  institution = {Advanced Photon Source},
  number = {LS-287},
  year = {2000},
  doi = {10.2172/761286}
}

@inproceedings{Pelegant,
  author = {Wang, Y. and Borland, M.},
  title = {Pelegant: A Parallel Accelerator Simulation Code for Electron Generation and Tracking},
  booktitle = {AIP Conference Proceedings},
  volume = {877},
  pages = {241--247},
  year = {2006},
  doi = {10.1063/1.2409141}
}

@article{chapman2023_fourthgen,
  author = {Chapman, Henry N.},
  title = {Fourth-generation light sources},
  journal = {IUCrJ},
  volume = {10},
  pages = {246--247},
  year = {2023},
  doi = {10.1107/S2052252523003585}
}

@article{raimondi2023_ebs,
  author = {Raimondi, Pantaleo and Benabderrahmane, Chamseddine and Berkvens, Paul and others},
  title = {The {Extremely Brilliant Source} storage ring of the {European Synchrotron Radiation Facility}},
  journal = {Communications Physics},
  volume = {6},
  pages = {82},
  year = {2023},
  doi = {10.1038/s42005-023-01195-z}
}

@article{jiao2026_heps_commissioning,
  author = {Jiao, Yi and Li, Ming and He, Ping and others},
  title = {Design and commissioning of the {High Energy Photon Source}},
  journal = {AAPPS Bulletin},
  volume = {36},
  pages = {3},
  year = {2026},
  doi = {10.1007/s43673-026-00184-y}
}

@article{peng2020_heps_booster,
  author = {Peng, Yuemei and Duan, Zhe and Guo, Yuanyuan and others},
  title = {Design of the {HEPS} booster lattice},
  journal = {Radiation Detection Technology and Methods},
  volume = {4},
  pages = {425--432},
  year = {2020},
  doi = {10.1007/s41605-020-00202-z}
}

@techreport{fornek2019_apsu_fdr,
  author = {Fornek, Thomas E.},
  title = {{Advanced Photon Source Upgrade Project Final Design Report}},
  institution = {Argonne National Laboratory},
  year = {2019},
  month = {May},
  doi = {10.2172/1543138},
}

@inproceedings{harkay2019_par_highcharge,
  author = {Harkay, K. C. and Calvey, J. R. and others},
  title = {Circuit Model Analysis for High Charge in the {APS Particle Accumulator Ring}},
  booktitle = {Proceedings of NAPAC'19},
  venue = {Lansing, MI, USA},
  publisher = {JACoW Publishing, Geneva, Switzerland},
  pages = {151--154},
  year = {2019},
  doi = {10.18429/JACoW-NAPAC2019-MOPLM21},
}

@inproceedings{steier2018_alsu_concept,
  author = {Steier, C. and All{\'e}zy, A. P. and Anders, A. and others},
  title = {Status of the Conceptual Design of {ALS-U}},
  booktitle = {Proceedings of IPAC'18},
  venue = {Vancouver, Canada},
  publisher = {JACoW Publishing, Geneva, Switzerland},
  pages = {4134--4137},
  year = {2018},
  doi = {10.18429/JACoW-IPAC2018-THPMF036},
}

@inproceedings{calvey:2025:napac,
  author       = {J. Calvey and T. Berenc and J. Dooling and K. Harkay and O. Mohsen and A. Nassiri and A. Puttkammer and F. Rafael and H. Shang and T. Smith and Y. Sun and G. Waldschmidt and U. Wienands and K. P. Wootton},
  title        = {Operation of the {APS-U} injectors with high single bunch charge},
  booktitle    = {Proceedings of NAPAC'25},
  venue        = {Sacramento, CA, USA},
  year         = {2025},
  pages        = {20--25},
  paper        = {MOYN01},
  doi          = {10.18429/JACoW-NAPAC2025-MOYN01},
  isbn         = {978-3-95450-261-5},
  issn         = {2673-7000},
}

@article{zhang:2023:hepsrf,
  author         = {P. Zhang and J. Dai and Z. Deng and L. Guo and T. Huang and D. Li and J. Li and Z. Li and H. Lin and Y. Luo and Q. Ma and F. Meng and Z. Mi and Q. Wang and H. Xu and X. Zhang and F. Zhao and H. Zheng},
  title          = {Radio-frequency system of the {High Energy Photon Source}},
  journal        = {Radiation Detection Technology and Methods},
  volume         = {7},
  pages          = {159--170},
  year           = {2023},
  doi            = {10.1007/s41605-022-00366-w},
}

@article{wang2021_heps_5cell_kicker,
  author  = {Lei Wang and Jinhui Chen and Hua Shi and Guanwen Wang and Na Wang and Zhe Duan and Saike Tian and Lihua Huo and Peng Liu},
  title   = {A Novel 5-Cell Strip-Line Kicker Prototype for the {HEPS} On-Axis Injection System},
  journal = {Nuclear Instruments and Methods in Physics Research Section A},
  volume  = {992},
  pages   = {165040},
  year    = {2021},
  doi     = {10.1016/j.nima.2021.165040}
}

@article{liu:2021:heps_gts,
  author    = {Fang Liu and Ge Lei and Zhe Duan and Cai Meng and Chenyan Lu and Yanhua Shao and Xinpeng Ma and Nan Gan and Yuemei Peng and Jianshe Cao and Chungming Chu and Yanfeng Sui and Jingyi Li},
  title     = {The design of {HEPS} global timing system},
  journal   = {Radiation Detection Technology and Methods},
  volume    = {5},
  pages     = {379--388},
  year      = {2021},
  doi       = {10.1007/s41605-021-00257-6},
  issn      = {2509-9930}
}

@article{foosang:2025:prst-ab,
  author    = {W. Foosang and A. Gamelin and V. Gubaidulin and R. Nagaoka},
  title     = {Transverse beam instabilities in low-emittance booster synchrotrons},
  journal   = {Physical Review Accelerators and Beams},
  volume    = {28},
  pages     = {094402},
  year      = {2025},
  day       = {16},
  doi       = {10.1103/PhysRevAccelBeams.28.094402},
}

@article{Takaki2010PSM,
author  = {Takaki, H. and Nakamura, N. and Kobayashi, Y. and Harada, K. and Miyajima, T. and Ueda, A. and Nagahashi, S. and Shimada, M. and Obina, T. and Honda, T.},
title   = {Beam injection with a pulsed sextupole magnet in an electron storage ring},
journal = {Physical Review Special Topics -- Accelerators and Beams},
volume  = {13},
pages   = {020705},
year    = {2010},
doi     = {10.1103/PhysRevSTAB.13.020705}
}

@article{Leemann2012PulsedSextupole,
author  = {Leemann, S. C.},
title   = {Pulsed sextupole injection for {S}weden’s new light source {MAX IV}},
journal = {Physical Review Special Topics -- Accelerators and Beams},
volume  = {15},
pages   = {050705},
year    = {2012},
doi     = {10.1103/PhysRevSTAB.15.050705}
}

@article{Aiba2015LongitudinalInjection,
author  = {Aiba, M. and B{\"o}ge, M. and Marcellini, F. and Sa{\'a} Hern{\'a}ndez, A. and Streun, A.},
title   = {Longitudinal injection scheme using short pulse kicker for small aperture electron storage rings},
journal = {Physical Review Special Topics -- Accelerators and Beams},
volume  = {18},
pages   = {020701},
year    = {2015},
doi     = {10.1103/PhysRevSTAB.18.020701}
}

@inproceedings{Gough2017AntiSeptum,
author    = {Gough, C. and Aiba, M.},
title     = {Top-up injection with anti-septum},
booktitle = {Proceedings of IPAC'17},
pages     = {774--776},
year      = {2017},
doi       = {10.18429/JACoW-IPAC2017-MOPIK104}
}

@article{Ehrlichman2021ALSUKicker,
author  = {Ehrlichman, M. and Hellert, T. and Leemann, S. C. and Penn, G. and Steier, C. and Sun, C. and Venturini, M. and Wang, D.},
title   = {Three-dipole kicker injection scheme for the {Advanced Light Source} upgrade accumulator ring},
journal = {Physical Review Accelerators and Beams},
volume  = {24},
pages   = {120702},
year    = {2021},
doi     = {10.1103/PhysRevAccelBeams.24.120702}
}

@article{CEPCAcceleratorTDR,
  author  = {{CEPC Study Group}},
  title   = {{CEPC Technical Design Report: Accelerator}},
  journal = {Radiation Detection Technology and Methods},
  volume  = {8},
  pages   = {1--1105},
  year    = {2024},
  doi     = {10.1007/s41605-024-00463-y},
  eprint  = {2312.14363},
  archivePrefix = {arXiv},
  primaryClass  = {physics.acc-ph}
}

@article{Harada2007PulsedQuadrupole,
  author  = {Harada, K. and Kobayashi, Y. and Miyajima, T. and Nagahashi, S.},
  title   = {New injection scheme using a pulsed quadrupole magnet in electron storage rings},
  journal = {Physical Review Special Topics -- Accelerators and Beams},
  volume  = {10},
  pages   = {123501},
  year    = {2007},
  doi     = {10.1103/PhysRevSTAB.10.123501}
}

@article{Alexandre2021MIK,
  author  = {Alexandre, Patrick and Ben El Fekih, Rachid and Letr{\'e}sor, Antoine and Thoraud, Serge and da Silva Castro, Jos{\'e} and Bouvet, Fran{\c c}ois and Breunlin, Jonas and Andersson, {\AA}ke and Tavares, Pedro Fernandes},
  title   = {Transparent top-up injection into a fourth-generation storage ring},
  journal = {Nuclear Instruments and Methods in Physics Research Section A},
  volume  = {986},
  pages   = {164739},
  year    = {2021},
  doi     = {10.1016/j.nima.2020.164739}
}

@article{Ollier2023MIK,
  author  = {Ollier, R. and Alexandre, P. and Ben El Fekih, R. and Breunlin, J. and da Silva Castro, J. and Thoraud, S. and Andersson, {\AA}. and Tavares, P. F.},
  title   = {Toward transparent injection with a multipole injection kicker in a storage ring},
  journal = {Physical Review Accelerators and Beams},
  volume  = {26},
  pages   = {020101},
  year    = {2023},
  doi     = {10.1103/PhysRevAccelBeams.26.020101}
}

@article{Borland2014DLSR,

  author    = {Michael Borland and Glenn Decker and Louis Emery and Vadim Sajaev and Yipeng Sun and Aimin Xiao},

  title     = {Lattice design challenges for fourth-generation storage-ring light sources},

  journal   = {Journal of Synchrotron Radiation},

  year      = {2014},

  volume    = {21},

  pages     = {912--936},


  doi       = {10.1107/S1600577514015203},

  issn      = {1600-5775}

}

@article{Tavares2018MAXIV,
  author  = {Tavares, Pedro F. and Al-Dmour, Eshraq and Andersson, {Å}ke and Cullinan, Fredrik and Jensen, B. N. and Olsson, D. and Olsson, D. K. and Sj{"o}str{"o}m, M. and Tarawneh, H. and Thorin, S. and Vorozhtsov, A.},
  title   = {Commissioning and first-year operational results of the {MAX IV} 3 {GeV} ring},
  journal = {Journal of Synchrotron Radiation},
  volume  = {25},
  pages   = {1291--1316},
  year    = {2018},
  doi     = {10.1107/S1600577518008111}
}

@inproceedings{Liu2021Sirius,
  author    = {Liu, L. and Alves, M. B. and Oliveira, A. C. S. and Resende, X. R. and de S{\'a}, F. H.},
  title     = {{Sirius} commissioning results and operation status},
  booktitle = {Proceedings of IPAC'21},
  publisher = {JACoW Publishing, Geneva, Switzerland},
  pages     = {13--18},
  year      = {2021},
  doi       = {10.18429/JACoW-IPAC2021-MOXA03}
}

@inproceedings{Sajaev2025APSU,
  author    = {Sajaev, V. and Borland, M. and Calvey, J. and Dooling, J. and Emery, L. and Harkay, K. and Kuklev, N. and Mohsen, O. and Sun, Y. and Arnold, N. and Berenc, T. and Brill, A. and Bui, H. and Carwardine, J. and Cheng, W. and Fors, T. and Kelly, M. and Lindberg, R. and Shen, G. and Smith, M. and Rafael, F. and Shang, H. and Soliday, R. and Wienands, U. and Yang, B.},
  title     = {{APS} upgrade: Commissioning the world's first light source based on swap-out injection},
  booktitle = {Proceedings of NAPAC'25},
  publisher = {JACoW Publishing, Geneva, Switzerland},
  year      = {2025},
  doi       = {10.18429/JACoW-NAPAC2025-MOXP02}
}

@article{Aiba2026Crosstalk,
  author  = {Aiba, M. and B{"o}ge, M. and Braun, H. and Calzolaio, C. and Dreyer, K. and Ganter, R. and Lerch, P. and Montenero, G. and Riccioli, R. and Riemann, B. and Sanfilippo, S. and Sidorov, S. and Streun, A. and Vrankovi{\'c}, V. and Zoller, C.},
  title   = {Realization of compact permanent-magnet-based multibend-achromat lattice with magnetic crosstalk compensations},
  journal = {Physical Review Accelerators and Beams},
  volume  = {29},
  pages   = {022402},
  year    = {2026},
  doi     = {10.1103/tfjh-x49q}
}

@article{Schroer2018PETRAIV,
  author  = {Schroer, C. G. and Agapov, I. and Brefeld, W. and others},
  title   = {{PETRA IV}: the ultralow-emittance source project at {DESY}},
  journal = {Journal of Synchrotron Radiation},
  volume  = {25},
  pages   = {1277--1290},
  year    = {2018},
  doi     = {10.1107/S1600577518008858}
}

@article{Grishanov1997FastKicker,
  author  = {Grishanov, B. I. and Podgorny, F. V. and R{"u}mmler, J. and Shiltsev, V. D.},
  title   = {Very fast kicker with high repetition rate for accelerator applications},
  journal = {Nuclear Instruments and Methods in Physics Research Section A},
  volume  = {396},
  pages   = {28--34},
  year    = {1997},
  doi     = {10.1016/S0168-9002(97)00739-0}
}

@article{Sajaev2019Collective,
  author  = {Sajaev, V. and Lindberg, R. R. and Borland, M. and Shin, S.},
  title   = {Simulations and measurements of the impact of collective effects on dynamic aperture},
  journal = {Physical Review Accelerators and Beams},
  volume  = {22},
  pages   = {032802},
  year    = {2019},
  doi     = {10.1103/PhysRevAccelBeams.22.032802}
}

@inproceedings{Lindberg2019SwapOut,
  author    = {Lindberg, R. R.},
  title     = {Controlling transient collective instabilities during swap-out injection},
  booktitle = {Proceedings of NAPAC'19},
  publisher = {JACoW Publishing, Geneva, Switzerland},
  pages     = {110--112},
  year      = {2019},
  doi       = {10.18429/JACoW-NAPAC2019-MOPLM08}
}

@inproceedings{Calvey2021APSBooster,
  author    = {Calvey, J. R. and Dooling, J. C. and Harkay, K. C. and Wootton, K. P. and Yao, C.},
  title     = {Measurements and simulations of high charge beam in the {APS} booster},
  booktitle = {Proceedings of IPAC'21},
  publisher = {JACoW Publishing, Geneva, Switzerland},
  pages     = {197--200},
  year      = {2021},
  doi       = {10.18429/JACoW-IPAC2021-MOPAB045}
}

\end{document}